\documentclass[aps,pra,epsfigure,twocolumn]{revtex4}
\usepackage{dcolumn}    
\usepackage{bm,lettrine} 
\usepackage{graphicx}
\usepackage{amsmath}    
\usepackage{latexsym}
\usepackage{amsfonts}   
\usepackage{amssymb}
\usepackage{array}      
\usepackage{epsfig}
\usepackage{txfonts}
\usepackage{color}
\usepackage[colorlinks=true,linkcolor=blue,urlcolor=blue,citecolor=blue]{hyperref}
\usepackage{hyperref}

\newcommand{\ket}[1]{\left\vert#1\right\rangle}

\newcommand{\nbar}{\overline{n}}

\begin{document}
\title{Equilibration and nonclassicality of a double-well potential}
       
\author{Steve Campbell}
\affiliation{Centre for Theoretical Atomic, Molecular and Optical Physics, Queen's University Belfast, Belfast BT7 1NN, United Kingdom}
\author{Gabriele De Chiara}
\affiliation{Centre for Theoretical Atomic, Molecular and Optical Physics, Queen's University Belfast, Belfast BT7 1NN, United Kingdom}
\author{Mauro Paternostro}
\affiliation{Centre for Theoretical Atomic, Molecular and Optical Physics, Queen's University Belfast, Belfast BT7 1NN, United Kingdom}

\begin{abstract}
A double well loaded with bosonic atoms represents an ideal candidate to simulate some of the most interesting aspects in the phenomenology of thermalisation and equilibration. Here we report an exhaustive analysis of the dynamics and steady state properties of such a system locally in contact with different temperature reservoirs. We show that thermalisation only occurs `accidentally'. We further examine the nonclassical features and energy fluxes implied by the dynamics of the double-well system, thus exploring its finite-time thermodynamics in relation to the settlement of nonclassical correlations between the wells.
\end{abstract}
\date{\today}
\maketitle

The high degree of control available when dealing with ultracold atomic samples makes them ideal candidates for realising prototypical quantum technology devices~\cite{Sanpera2012,review_optical_lattices}. The range of practical applications that can be addressed using platforms based on the physics of ultracold atomic ensembles ranges from metrology and sensing to the achievement of quantum memories~\cite{brennen}, from ultra-stable atomic clocks~\cite{andre} to the simulation of difficult condensed-matter physics problems~\cite{georgescu}. Recently, such range has been extended to quantum thermometry~\cite{mehboudi}, while theoretical and experimental interest is emerging in the design and implementation of thermodynamic processes and (elementary) engines based on such systems~\cite{brantut}. The tuneable interactions among the elementary constituents of a cold-atom system, and  the availability of effective ways of arranging non-equilibrium states of atomic systems confined in external optical potentials, provide an almost ideal scenario for the study and harnessing of thermodynamically relevant questions and tasks, indeed recently thermal and number fluctuations have been studied for ultracold atoms in two mode traps~\cite{bruno}.

For such endeavours to succeed, it is absolutely crucial to identify a suitable configuration to act as the basic building block for a thermodynamic device, and characterise its working principles in terms of fundamental quantities (such as heat and work), which will pave the way to the actual construction of the machine itself. 

In this paper, we move exactly along these lines: Inspired by the experimental set up of Refs.~\cite{brantut}, where a cold atomic system is placed in contact with two different thermal reservoirs, we consider a slight modification in which the gate potential separating the two reservoirs is replaced by a double well potential loaded with a Bose-Einstein condensate (BEC), itself a system of vast experimental implementability~\cite{smerzi,albiez}. We set and study explicitly its non-equilibrium dynamics. By assuming each well is initially thermalised to its own local reservoir, we will show that, in the tunnelling dominated regime, a temperature imbalance between the wells  leads to the emergence of non-classicality, and study how this is linked to the equilibration dynamics of the atomic system. Remarkably, we show that the genuinely quantum nature of the state of the double well does not appear to affect the rate of equilibration of the open system at hand. By working in the weak coupling regime between each well and its reservoir, which allows us to identify clearly the contributions of each well to the total heat flux into/out of the local environments, we highlight a rather rich and complex dynamics of the heat exchange across the wells. Further, we examine its relation with the emergence of nonclassical correlations within the state of the atomic ensemble within a vast range of operating conditions. 

\section{Description of the model}
We are interested in studying the out of equilibrium dynamics and steady-state properties of a system of cold atoms loaded in a double-well potential and subject to the effects of two reservoirs at different energy. Our Hamiltonian is the two-site Bose-Hubbard model~\cite{smerzi} given by $\hat{\cal H}=\hat{\cal H}_f+\hat{\cal H}_{si}+\hat{\cal H}_{t}$ with [we assume units such that $\hbar=1$ throughout]
\begin{equation}
\label{hamiltonian}
\begin{aligned}
&\hat{\cal H}_f=\omega_1 \left(\hat a^\dagger_1 \hat a_1+\frac{1}{2} \right) + \omega_2 \left(\hat a^\dagger_2 \hat a_2+\frac{1}{2} \right),\\
&\quad\hat{\cal H}_{si}= \frac{U}{2} \left(\hat a^{\dagger2}_1\hat a^2_1+\hat a^{\dagger2}_2\hat a^2_2 \right),\\
&\quad\hat{\cal H}_t= - {\cal J} \left(\hat a^\dagger_1 \hat a_2 + \hat a_1 \hat a^\dagger_2\right).
\end{aligned}
\end{equation}
Here $\hat{\cal H}_f$ describes the free evolution of the atomic systems in the two wells, each occurring at the rate set by the single-atom energy $\omega_j$, and with $\hat a_j,~\hat a_j^\dagger$ the associated annihilation and creation operators for each well. The Hamiltonian term $\hat{\cal H}_{si}$ accounts for the self-interaction (at rate $U$) between atoms occupying the same well, while $\hat{\cal H}_{t}$ stands for the tunnelling term, which occurs at rate ${\cal J}$. We will focus mostly on the tunnelling-dominated regime associated with $U=0$. However, the interaction-dominated regime corresponding to ${\cal J}=0$, and the intermediate regime will also be addressed. The focus of our investigation will be the phenomenology of thermalisation of the system, both at the single and two-well level. We remark that model, Eq.~\eqref{hamiltonian}, can be realised in a variety of settings including superconducting Josephson junctions~\cite{schon}, trapped ions~\cite{ions}, bimodal optical cavities~\cite{larson} and optomechanical setups~\cite{markus}.

While important insight will be gathered by addressing the unitary evolution induced by considering $\hat{\cal H}$, the overarching goal of this work is the study of the open-system evolution created by the contact of the two wells with their respective reservoirs. We are interested in addressing the dynamics induced by the master equation~\cite{breuer}
\begin{equation}
\label{master}
\dot{\varrho}_t=-i\left[\hat{\cal H},\varrho_t  \right] + \sum_{j=1}^2 \mathcal{L}_j(\varrho_t),
\end{equation}
where we have introduced the overall-system density matrix at a generic time $t$, $\varrho_t$, and the Lindblad super-operators
\begin{equation}
\begin{aligned}
\mathcal{L}_j (\varrho_t)=& \gamma_j (\nbar_j+1)\left( \hat a_j \varrho_t \hat a_j^\dagger -\frac{1}{2} \{\hat a_j^\dagger \hat a_j, \varrho_t\}  \right) +\\
                                         &  \gamma_j \nbar_j\left( \hat a_j^\dagger \varrho_t \hat a_j -\frac{1}{2} \{\hat a_j \hat a_j^\dagger, \varrho_t\} \right), 
\end{aligned}
\end{equation}
which describe the incoherent particle-exchange process (occurring at rate $\gamma_j$) between a well and the respective reservoir (assumed to have a thermal occupation number $\nbar_j$). Eq.~\eqref{master} is the key equation in our analysis to follow. We remark that in certain working conditions this description of the dynamics is not always valid. In particular, when the scattering length of the BEC is large, non-Markovian dynamics can play an important role~\cite{haikka}. We therefore assume that the scattering length is sufficiently small to ensure the validity of the Markovian approximation~\cite{haikka}.\\
\\
\section{Exact solutions of the tunneling-dominated regime.}
In order to gather insight into the basic coherent processes of the system in the case of tunnelling-dominated regimes, we set $U=0$ in $\hat{\cal H}$ and address the unitary evolution first. We define the canonical quadrature operators $\{\hat x_1,\hat p_1,\hat x_2,\hat p_2\}$ as~\cite{carmicheal,walls}
\begin{equation}
\hat x_j=\frac{1}{\sqrt{2}}\left( \hat a_j + \hat a^\dag_j  \right),~~~\text{and}~~~\hat p^\dagger_j=\frac{i}{\sqrt{2}}\left( \hat a^\dag_j - \hat a_j  \right),
\end{equation}
and recast the Hamiltonian into the form
\begin{equation}
\label{withquad}
\hat {\cal H}=\frac{\omega_1}{2}\left( \hat x_1^2 + \hat p_1^2 + \openone \right) + \frac{\omega_2}{2}\left( \hat x_2^2 + \hat p_2^2 + \openone \right) - {\cal J} \left( \hat x_1 \hat x_2 + \hat p_1 \hat p_2 \right),
\end{equation}
with $\openone$ the identity operator. By neglecting trivial constant terms, Eq.~\eqref{withquad} can be thus interpreted as a quadratic form identified by the adjacency matrix
\begin{equation}
{\cal A}=
\left( \begin{array}{cccc}
\omega_1 & 0 & -{\cal J} & 0 \\
0 & \omega_1 & 0 & -{\cal J} \\
-{\cal J} & 0 & \omega_2 & 0  \\
 0 & -{\cal J} & 0 & \omega_2  \\
\end{array}
\right),
\end{equation}
which has been written in the ordered operator basis $\{\hat x_1,\hat p_1,\hat x_2,\hat p_2\}$. In what follows, we rescale all the relevant frequencies with respect to $\omega_1$. In these units, we have $\omega_2\rightarrow\omega_2/\omega_1=1+\Delta$ with $\Delta$ a dimensionless bias between the two wells, and ${\cal J}\to J={\cal J}/\omega_1$. The rescaled Hamiltonian $\hat{\cal H}/\omega_1$ can be diagonalised by means of a simple two-mode mixing transformation $\hat U_T(\theta)=\exp[-i\tfrac{\theta}{2}(\hat a^\dag_1\hat a_2+\hat a_1 \hat a^\dag_2)]$ with $\theta=-\frac{1}{2}\arctan\left({2J}/{\Delta}\right)$, which leaves us with the new quadratic Hamiltonian
\begin{equation}
{{\hat{\cal H}}_q}/\omega_1=\Omega_1 ( \hat{x}_1^2 + \hat{p}_1^2  ) + \Omega_2 ( \hat{x}_2^2 + \hat{p}_2^2  ), 
\end{equation}
describing two freely evolving harmonic oscillators at the respective frequencies $\Omega_1=1+({\Delta - \Gamma})/{2},~~\Omega_2=1+({\Delta + \Gamma})/{2}$, with $\Gamma=\sqrt{\Delta^2+4 J^2}$. For a Gaussian initial state of the system~\cite{walls}, given the quadratic nature of the Hamiltonian, rather than tracking the evolution of the density matrix of the system, we can restrict our attention to the evolved form of the covariance matrix $\sigma$ of entries $\sigma_{ij}=\langle\{\hat P_i,\hat P_j\}\rangle-\langle \hat P_i\rangle\langle \hat P_j\rangle$, where $\hat P_i$'s are the elements of the vector of quadrature operators $\hat P^\top=(\hat x_1~\hat p_1~\hat x_2~\hat p_2)$ and the expectation value of such vector (calculated over the state of the system), which bear full information on the state of the system. Both are readily gathered as  
\begin{equation}
\label{evolv}
\sigma_u(t)=M \sigma_u(0) M^\top,\qquad\langle{\hat P}\rangle_t=M \langle{\hat P}\rangle_0,
\end{equation}
with $M=T(\theta) R_1(\Omega_1 t)R_2 (\Omega_2 t) T(\theta)^\top$, $\sigma_u(0)$ [$\langle{\hat P}\rangle_0$] the covariance matrix [vector of phase-space displacements] of the initial state of the system and $\sigma_u(t)$ [$\langle{\hat P}\rangle_t$] its time-evolved version. In Eq.~\eqref{evolv} $R_j(\Omega_j t)~(j=1,2)$ and $T(\theta)$ are the symplectic transformations corresponding to the free evolution $e^{-i\Omega_j\hat a^\dag_j\hat a_j t}$ and two-mode mixing $\hat U_T(\theta)$. Explicitly
\begin{equation}
\begin{aligned}
& R_j(\Omega_jt)=
\begin{pmatrix}
\cos(\Omega_j t)&\sin(\Omega_j t)\\
-\sin(\Omega_j t)&\cos(\Omega_j t)
\end{pmatrix},\\
&
T(\theta)=\left( \begin{array}{cccc}
\cos\theta & 0 & \sin\theta & 0 \\
0 & \cos\theta & 0 & \sin\theta  \\
-\sin\theta  & 0 & \cos\theta & 0  \\
 0 & -\sin\theta  & 0 & \cos\theta  \\
\end{array}
\right).
\end{aligned}
\end{equation}
We now concentrate on the situation where the particles in each well are initially at thermal equilibrium with their local reservoirs. The initial covariance matrix will thus be that of a two-mode thermal state
\begin{equation}
\label{initial}
\sigma_u(0) = \left( \begin{array}{cccc}
2\nbar_1+1 & 0 & 0 & 0 \\
0 & 2\nbar_1+1  & 0 & 0 \\
0 & 0 & 2\nbar_2+1 & 0  \\
 0 & 0 & 0 & 2\nbar_2+1   \\
\end{array}
\right),
\end{equation}
with $\nbar_j=\langle\hat a^\dag_j\hat a_j\rangle$ the mean number of particles in the $j^{\rm th}$ well. For such an initial state, the phase-space displacements are all null and full information on the evolved state is provided by the covariance matrix
\begin{equation}
\label{unitarysol}
\sigma_u(t) = \left( \begin{array}{cccc}
n_1  & 0 & c_1 & c_2 \\
0 & n_1 &  -c_2 & c_1  \\
c_1 & -c_2  & n_2   & 0  \\
c_2  & c_1  & 0 &  n_2   \\
\end{array}
\right),
\end{equation}
with elements
\begin{equation}
\begin{aligned}
&c_1=\frac{4J\Delta\left( \nbar_1-\nbar_2 \right)\sin\left(\Gamma t/2\right)}{\Gamma^2},~c_2=\frac{2{J} \left( \nbar_1 - \nbar_2 \right) \sin\left(\Gamma t\right)}{\Gamma},\\
&n_1=\frac{{4 J^2 (\nbar_1-\nbar_2) \cos \left(\Gamma t \right)+4J^2 (\nbar_1+\nbar_2+1)+\Delta ^2 (2 \nbar_1+1)}}{\Gamma^2},\\
&n_2=\frac{{4 J^2 (\nbar_2-\nbar_1) \cos \left(\Gamma t \right)+4J^2 (\nbar_1+\nbar_2+1)+\Delta ^2 (2 \nbar_2+1)}}{\Gamma^2}.
\end{aligned}
\end{equation}
If both wells are at the same initial temperature, i.e. $\nbar_1=\nbar_2$, then $c_1=c_2=0$ and $n_1=n_2=2\nbar_1+1$, i.e. the system does not evolve in time and the two wells remain at their thermal equilibrium, notwithstanding the tunneling. This is a clear interference effect. Moreover, for identical single-atom energy in each well, i.e. $\Delta=0$, $c_1$ is null, showing that the position [momentum] $\hat x_1$ [$\hat p_1$] gets correlated with $\hat p_2$ [$\hat x_2$]. In general, such correlations do not imply necessarily the setting of entanglement between the wells~\cite{Ferraro}. Indeed, the tunneling term of the Hamiltonian, $\hat {\cal H}_{t}$ in Eq.~\eqref{hamiltonian}, can generate entanglement only when the state of at least one of the wells is sufficiently non-classical. In the context of our investigation here, this basically implies the preparation of squeezed states of the wells~\cite{asboth}. This can be understood by noticing the formal analogy between $\hat {\cal H}_{t}$ and the generator of a two-mode mixing transformation and considering, for the sake of argument, the resonant case $\Delta=0$. Under such conditions, moving to the interaction picture with respect to $\hat{\cal H}_{f}$, the time evolution operator would correspond to $\hat U_T(Jt)$, which gives rise to no entanglement between the two wells when they are prepared in thermal states (even at different effective temperatures), as demonstrated in Ref.~\cite{asboth}. However, this does not imply that the dynamics of the two-well system is trivial. In fact, in general, quantum correlations (of a form weaker than entanglement) are set by $\hat U(Jt)$ when acting on thermal states with $\nbar_1\neq\nbar_2$. We will address the emergence of discord-like quantum correlations~\cite{zurek,vedral,discordreview} and its relation to the inter-well exchange process in a later section. 

We now move to solving the full dissipative dynamics governed by Eq.~\eqref{master} for $U=0$. The problem can  be efficiently solved by using a suitable Gaussian ansatz: We first translate Eq.~\eqref{master} into the phase space by deriving a differential equation for the symmetrically ordered characteristic function $\chi(\beta_1,\beta_2,t) = \text{Tr}[\hat{D}_1(\beta_1)\otimes \hat{D}_2(\beta_2)\varrho_t]$~\cite{carmicheal,walls}. Here $\hat D_j(\beta_j)=\exp[{\beta_j \hat a^\dagger_j - \beta_j^* \hat a_j}]$ is the Weyl displacement operator with amplitude $\beta_j\in{\mathbb C}$ for system $j=1,2$. Using the phase-space relations~\cite{walls}
\begin{equation}
\begin{split}
&\hat a^\dagger \hat D_j(\beta_j)\leftrightarrow\left( \frac{\partial}{\partial \beta} +\frac{\beta^*}{2} \right)\hat D_j(\beta_j),\\
&\hat D(\beta_j) \hat a^\dagger\leftrightarrow\left( \frac{\partial}{\partial \beta} - \frac{\beta^*}{2} \right)\hat D_j(\beta_j),\\
&\hat a \hat D_j(\beta_j)\leftrightarrow\left(\frac{\beta}{2} - \frac{\partial}{\partial \beta^*}\right)\hat D_j(\beta_j),\\
&\hat D_j(\beta_j) \hat a\leftrightarrow\left( - \frac{\beta}{2} - \frac{\partial}{\partial \beta^*} \right)\hat D_j(\beta_j),
\end{split}
\end{equation}
after a lengthy but otherwise straightforward calculation, we find that Eq.~\eqref{master} takes the form of the Fokker-Planck equation 
\begin{widetext}
\begin{equation}
\label{fulleq}
\begin{aligned}
\partial_t\chi(\beta_1,\beta_2)&=\left\{ iJ\left( -\beta_1\frac{\partial}{\partial\beta_2} -\beta_2\frac{\partial}{\partial\beta_1} + \beta_1^*\frac{\partial}{\partial\beta_2^*}+ \beta_2^*\frac{\partial}{\partial\beta_1^*} \right) \right.\\
&\left.- \sum_{j=1}^2\left[ \omega_j\left( \beta_j\frac{\partial}{\partial\beta_j}- \beta_j^*\frac{\partial}{\partial\beta_j^*} \right) + \frac{\gamma_j}{2}\left( \beta_j\frac{\partial}{\partial\beta_j} + \beta_j^*\frac{\partial}{\partial\beta_j^*}\right)
+ \gamma_j\left(\nbar_j+\frac12\right)\vert\beta_j\vert^2 \right] \right\} \chi(\beta_1,\beta_2).
\end{aligned}
\end{equation}
\end{widetext}
By letting $\beta_j=x_j+ip_j$ [so that $\chi(\beta_1,\beta_2,t)\to\chi(x_1,p_1,x_2,p_2,t)$] and expressing the characteristic function in terms of the entries of the vector of quadrature variables, we can write $\chi(x_1,p_1,x_2,p_2,t)=\exp[{iP^\top X - \tfrac{1}{2} P^\top \tilde\sigma P}]$, where we have introduced the generic vector of ${\mathbb C}$-numbers $X^\top=(y_1~z_1~y_2~z_2)$ and matrix $\tilde\sigma$ whose elements we aim at finding, which we do by solving Eq.~\eqref{fulleq}. In {\bf Methods} we provide the set of differential equations for the elements of $X$ and $\tilde\sigma$ obtained when evaluating both sides of Eq.~\eqref{fulleq} and equating them term by term. 

The explicit solution of the problem at hand leads to a time-evolved covariance matrix of the general block form
\begin{equation}
\label{explicit}
\sigma(t) =
\begin{pmatrix}
m_1\openone&{\bm c}\\
{\bm c}^\top&m_2\openone
\end{pmatrix},
\end{equation}
where ${\bm c}$ is a $2\times2$ matrix of correlations among the quadrature operators of the system. The diagonal structure of the blocks pertaining to the the individual wells shows that, locally, the system thermalises at temperatures determined by the explicit form of $m_{1,2}$. However, as ${\bm c}$ is, in general, not null, global thermalisation is not achieved: the  overall system never thermalises, notwithstanding an explicitly dissipative evolution. This is clearly seen by looking at the general form of the steady state. Although the analytic form of the non-zero elements is readily achievable for any value of the parameters involved in the problem, they are, in general, too cumbersome to be reported here.  However, assuming $\gamma_1=\gamma_2=\gamma$, the steady-state of the system is determined by the covariance matrix 
\begin{widetext}
\begin{equation}
\label{steadystate}
\sigma_{ss}=\zeta \left( \begin{array}{cccc}
 \tfrac{4J^2(\nbar_1+\nbar_2+1)}{\gamma^2+\Delta^2}+(2\nbar_1+1) & 0 & -\tfrac{2J\Delta(\nbar_1-\nbar_2)}{\gamma^2+\Delta^2} & \tfrac{2J\gamma\left( \nbar_1-\nbar_2 \right)}{\gamma^2+\Delta^2} \\
0 &  \tfrac{4J^2(\nbar_1+\nbar_2+1)}{\gamma^2+\Delta^2}+(2\nbar_1+1)  & -\tfrac{2J\gamma\left( \nbar_1-\nbar_2 \right)}{\gamma^2+\Delta^2}& -\tfrac{2J\Delta(\nbar_1-\nbar_2)}{\gamma^2+\Delta^2} \\
-\tfrac{2J\Delta(\nbar_1-\nbar_2)}{\gamma^2+\Delta^2} & -\tfrac{2J\gamma\left( \nbar_1-\nbar_2 \right)}{\gamma^2+\Delta^2} & \tfrac{4J^2(\nbar_1+\nbar_2+1)}{\gamma^2+\Delta^2}+(2\nbar_2+1)  & 0  \\
\tfrac{2J\gamma\left( \nbar_1-\nbar_2 \right)}{\gamma^2+\Delta^2} & -\tfrac{2J\Delta(\nbar_1-\nbar_2)}{\gamma^2+\Delta^2} & 0 & \tfrac{4J^2(\nbar_1+\nbar_2+1)}{\gamma^2+\Delta^2}+(2\nbar_2+1)  \\
\end{array}
\right),
\end{equation}
\end{widetext}
with $\zeta= \frac{\gamma^2+\Delta^2}{4 J^2+\gamma^2+\Delta^2}$. Clearly, only for $\nbar_1=\nbar_2$ the structure of the global covariance matrix takes a thermal-like form. However, this does not preclude the possibility to achieve accidental thermalisation, i.e. situations such that the state of the system either becomes globally/locally thermal, or closely approximates an equilibrium configuration. This will be the focus of the following analysis. \\

\begin{figure}[t!]
{\bf (a)} \hskip0.5\columnwidth {\bf (b)}\\
\includegraphics[width=0.5\columnwidth]{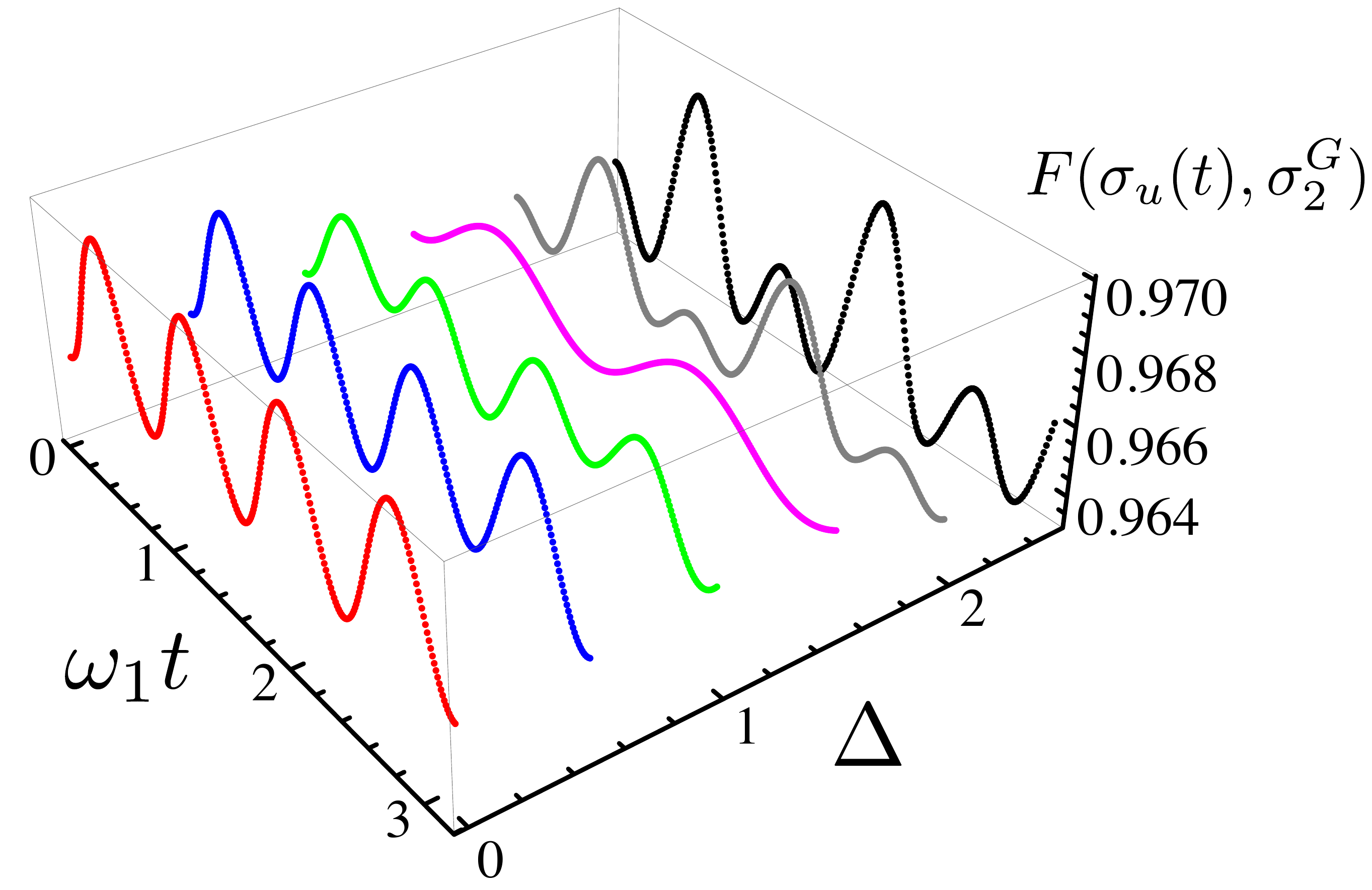}~\includegraphics[width=0.43\columnwidth]{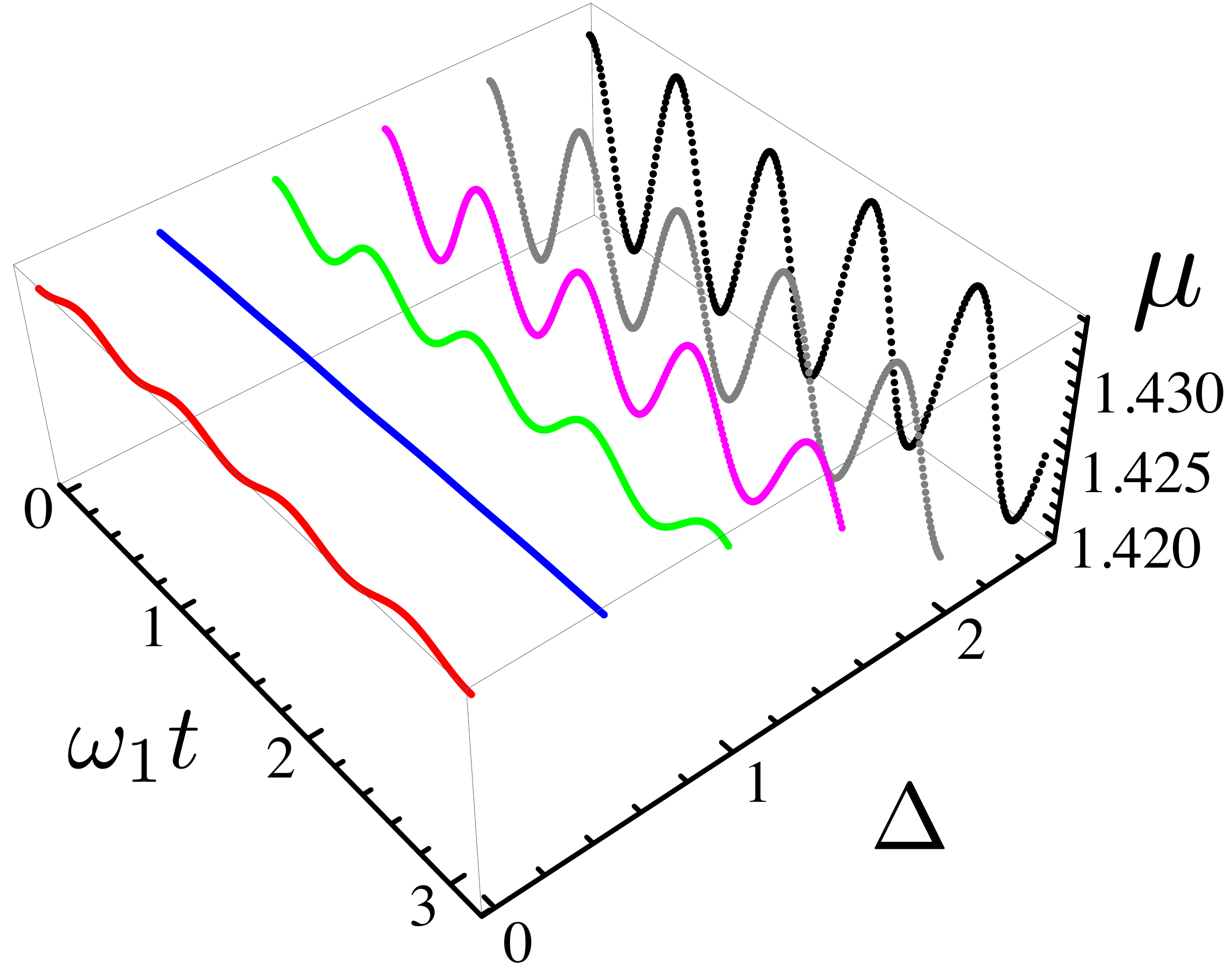}
\caption{{\bf (a)} Maximum fidelity between the instantaneous state of the system and a globally thermal state, plotted against the dimensionless evolution time and the (dimensionless) bias between the energies of the wells $\Delta$. {\bf (b)} Corresponding estimate of the mean energy $\mu$ of the target globally thermal state. In both panels, we have taken $J=2$, $\nbar_1=\nbar_2/2=1$.}
\label{fidelityunitaryglobal}
\end{figure}

\section{Assessment of dynamical thermalisation} 
We shall start with the study of the unitary case. Dynamical thermalisation in closed-system dynamics is a topic of vast interest, which has recently attracted considerable attention at both the theoretical and experimental level~\cite{Trotzky}. Our approach is based on the assessment of the distance between the time-dependent state of the system and a generic (either global or local) thermal state. Quantitatively, as a measure of the distance between two states $\rho_{1,2}$, we use the Ulhmann fidelity~\cite{uhlmann} 
\begin{equation}
F(\rho_1,\rho_2)=\left({\rm Tr}\sqrt{\sqrt{\rho_1}\rho_2\sqrt{\rho_1}}\right)^2.
\end{equation}
For Gaussian states, it can be conveniently evaluated using the  covariance matrices $\sigma_{1,2}$ associated to the states under scrutiny. The explicit formula, which has been recently reported in Ref.~\cite{marian}, reads
\begin{equation}
F(\sigma_1,\sigma_2)=4\frac{(\sqrt{x}+\sqrt{x-1})^2}{\sqrt{{\rm det}(\sigma_1+\sigma_1)}},
\end{equation}
where $\Omega=i(\sigma_y\oplus\sigma_y)$ is the two-mode symplectic matrix ($\sigma_y$ being the $y$-Pauli matrix) and $x=2\sqrt{{\cal I}_1}+2\sqrt{{\cal I}_2}+1/2$ with
\begin{equation}
{\cal I}_1=\frac{{\rm det}(\Omega\sigma_1\Omega\sigma_1-\openone_4)}{16{\rm det}(\sigma_1+\sigma_2)},~~{\cal I}_2=\frac{{\rm det}(\sigma_1+i\Omega){\rm det}(\sigma_2+i\Omega)}{16{\rm det}(\sigma_1+\sigma_2)}.
\end{equation}
In our case we consider $\sigma_1=\sigma_u(t)$ [cf. Eq.~\eqref{unitarysol}] and $\sigma_2$ given by either $\sigma^G_2=(2\mu+1)\openone_4$, i.e. the covariance matrix  of a globally thermal state with mean number of excitations $\mu$, or $\sigma^L_2=(2\mu_1+1)\openone_2\oplus(2\mu_2+1)\openone_2$, which is the one associated with the tensor product of locally thermal states (each with mean number of excitations $\mu_j$). For clarity, we have indicated with $\openone_n$ the identity matrix of dimension $n$. 

We present the case of global thermalisation first: after calculating the time behaviour of $F(\sigma_u(t),\sigma^G_2)$ for various choices of $\Delta$, we have numerically evaluated the value of $\mu$ that achieves the maximum of $F(\sigma_u(t),\sigma^G_2)$. In Fig.~\ref{fidelityunitaryglobal} we show both such value and the corresponding estimate for $\mu$. The state fidelity remains evidently quite large, being only partially depleted by an increasing value of $\Delta$ (the dependence on such parameter is quite non-trivial, given that for $\Delta=2.5$, for instance, values very close to those associated with $\Delta=0$ can be achieved, at suitable times in the evolution). However, while at small values of $\Delta$ the target state changes very little with time, this is not the case for increasing bias: the value of $\mu$ corresponding to a non-zero $\Delta$ oscillates with a non-negligible amplitude as this parameter grows. In any case, perfect thermalisation is never achieved, a result that is strengthened by the analysis that we will report in the next Section. 

\begin{figure}[t!]
{\bf (a)}\\
\includegraphics[width=0.5\columnwidth]{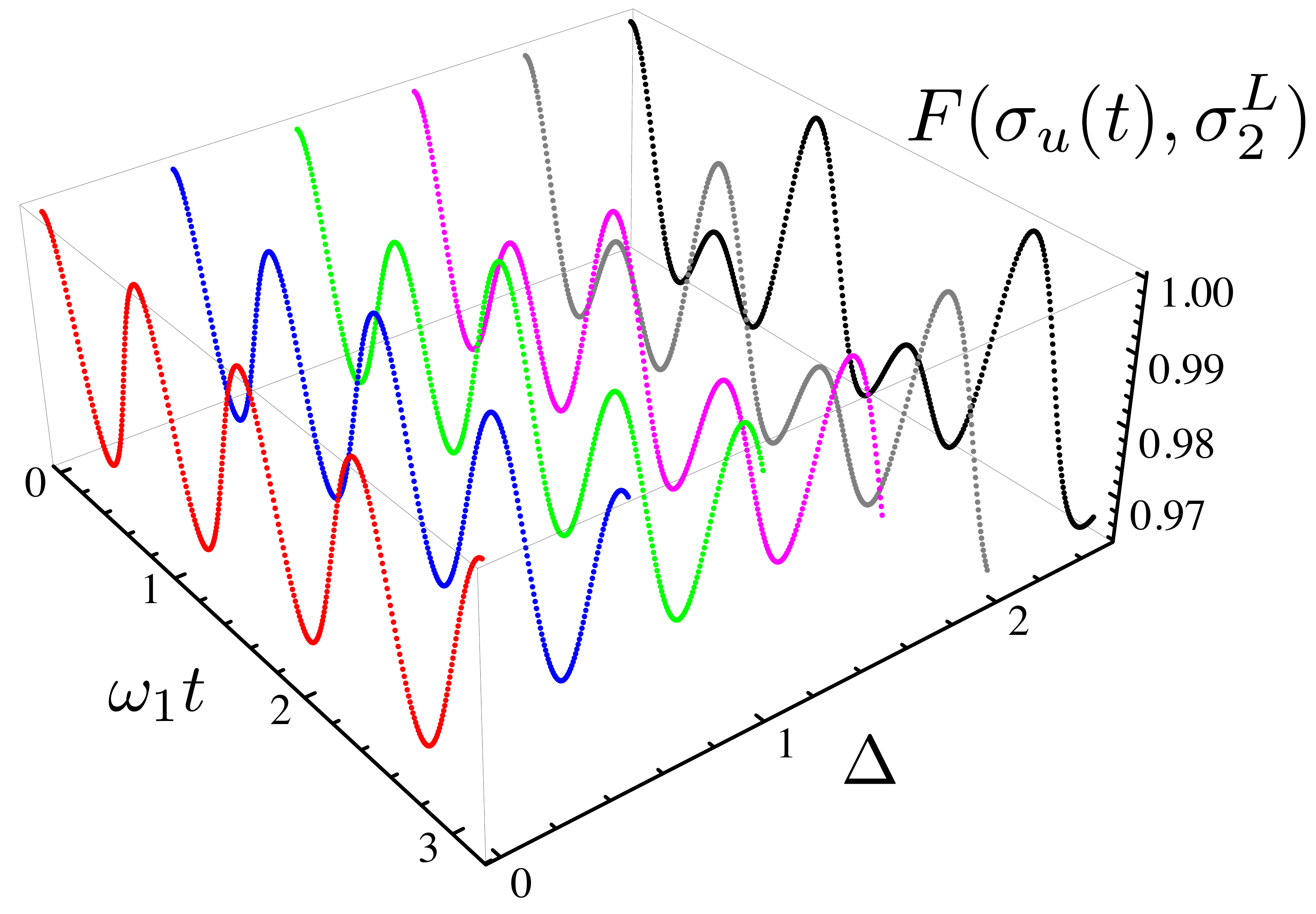}\\
{\bf (b)}\hskip0.33\columnwidth {\bf (c)}\\
\includegraphics[width=0.80\columnwidth]{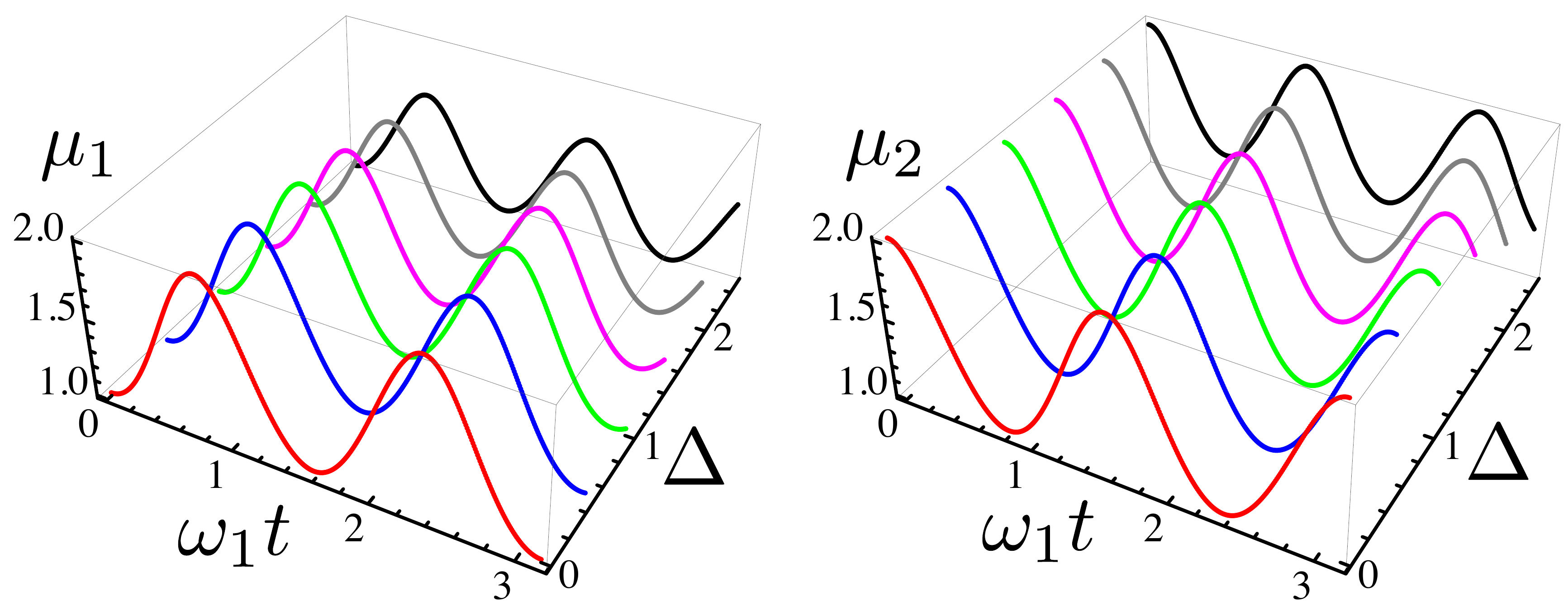}
\caption{{\bf (a)} Maximum fidelity between the instantaneous state of the system and a locally thermal state, plotted against the dimensionless evolution time and the (dimensionless) bias between the energies of the wells $\Delta$. {\bf (b)} \& {\bf (c)} Estimate of the corresponding mean number of excitations $\mu_{1,2}$ of the target locally thermal state. In both panels, we have taken $J=2$, $\nbar_1=\nbar_2/2=1$.}
\label{fidelityunitarylocal}
\end{figure}

The situation is somehow different when locally thermal target states are considered [cf. Fig.~\ref{fidelityunitarylocal}]: besides the expected times at which a full period of the evolution is achieved, it is possible to identify instants of time at which the state of the double-well system is indeed very close to a locally thermal state ($F(\sigma_u(t),\sigma^L_2)\ge0.999$), which would suggest the occurrence of accidental dynamical thermalisation.  

\begin{figure}[b!]
{\bf (a)}\\
\includegraphics[width=0.7\columnwidth]{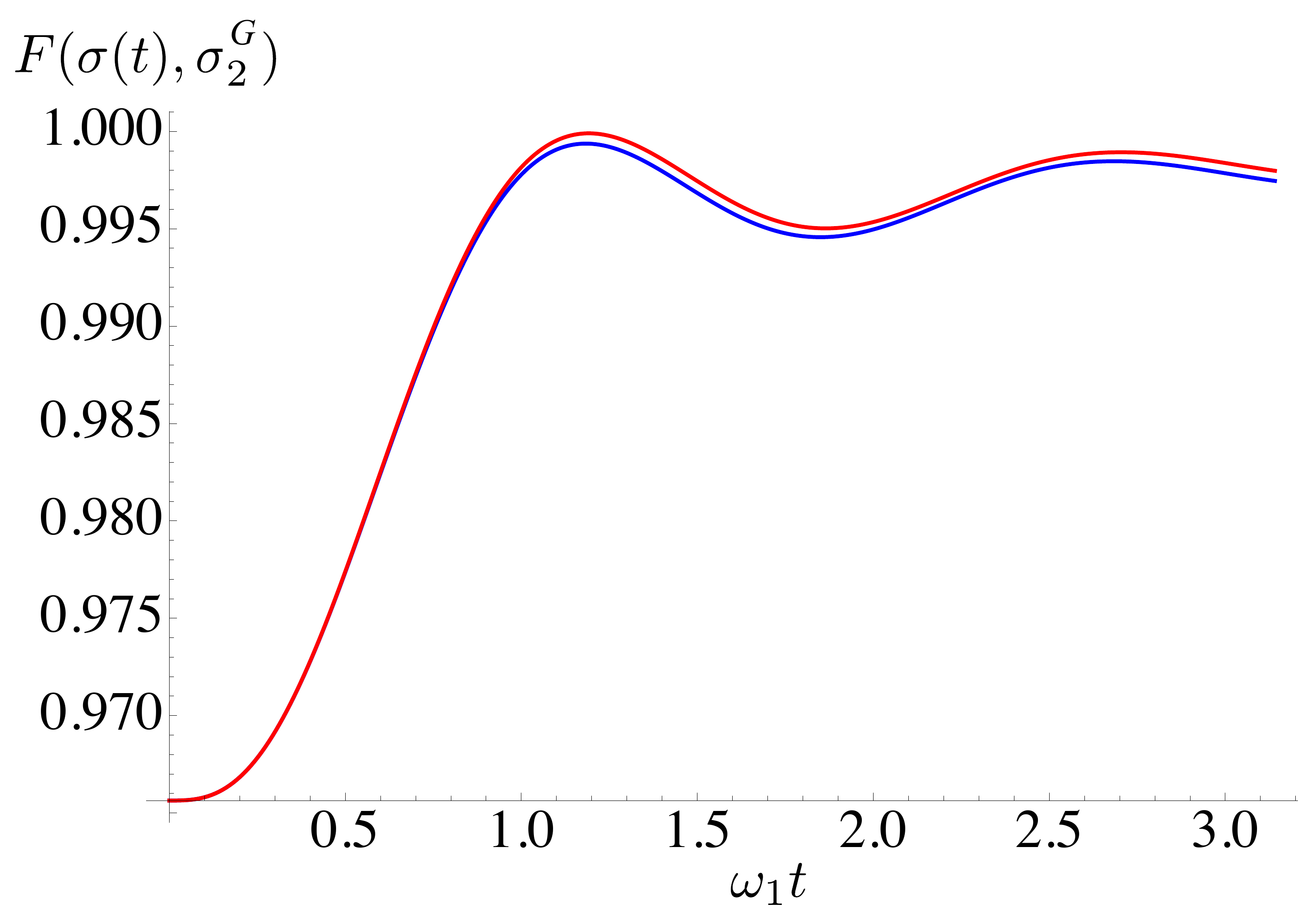}\\
{\bf (b)}\\
\includegraphics[width=0.7\columnwidth]{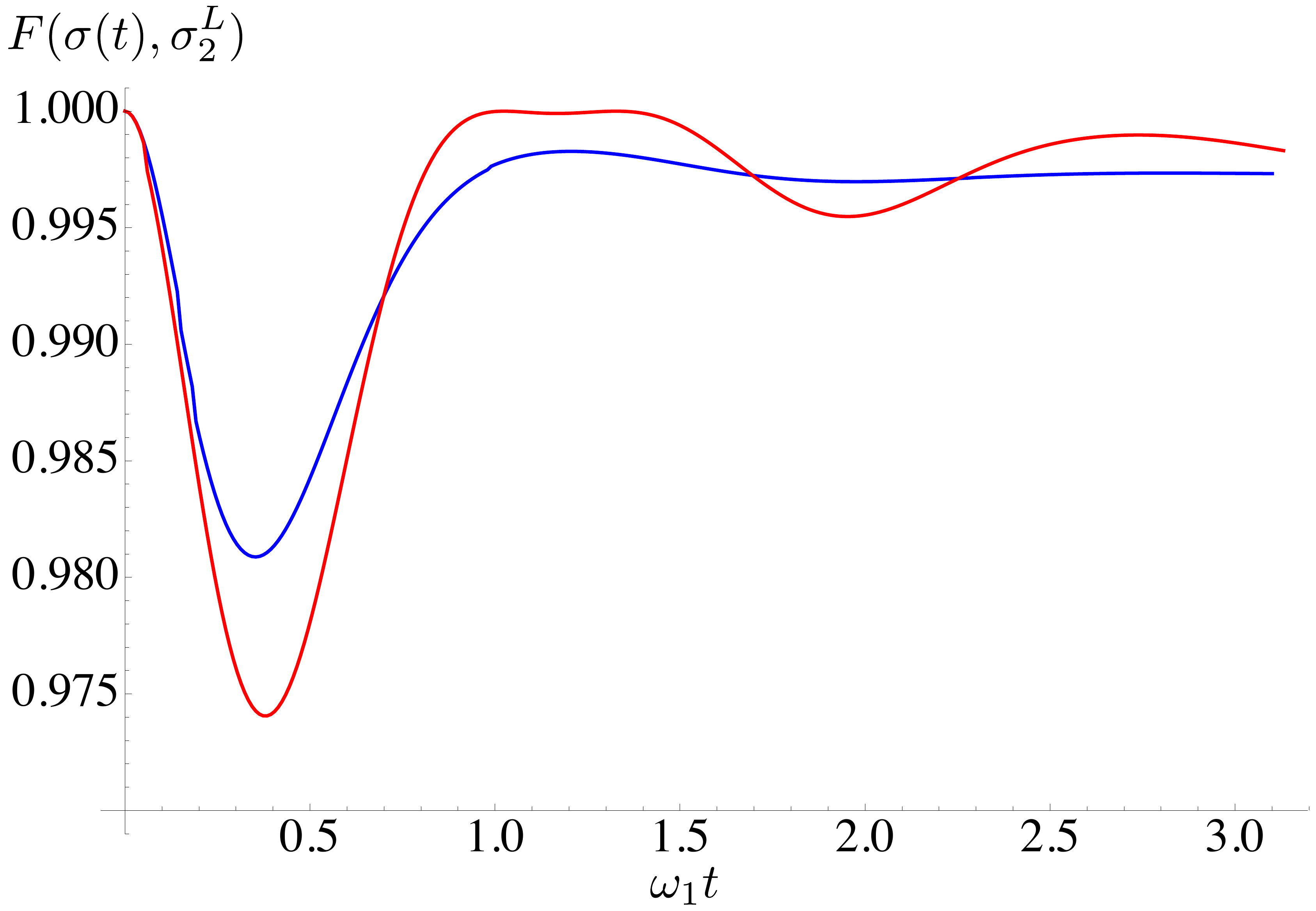}
\caption{Open-system dynamics. {\bf (a)} Fidelity with a globally thermal state for $J=2$, $\nbar_1=1$, $\nbar_2=2$, $\gamma_1/\omega_1=\gamma_2/\omega_1=1$. We have taken $\Delta=0$ (blue line) and $\Delta=0.5$ (red line). {\bf (b)} Fidelity with locally thermal states for $J=2$, $\nbar_1=1$, $\nbar_2=2$, and $\Delta=0$. We have taken $\gamma_1/\omega_1=\gamma_2/\omega_1=1$ (red line) and $\gamma_1/\omega_1=1$ with $\gamma_2/\omega_1=3$ (blue line).}
\label{fidelityopen}
\end{figure}
In the open-system dynamics case, a similar calculation allows us to evaluate the fidelity with both a globally thermal and locally thermal state as shown in Fig.~\ref{fidelityopen}, which studies the effects of both the energy bias [panel {\bf (a)}] and a difference in the damping rates of the two wells [panel {\bf (b)}]. Quite evidently, both effects spoil the state fidelity, which however achieves values that are either precisely 1 or very close to it. Indeed, focusing on the unbiased case with $\Delta=0$ and $\gamma_1=\gamma_2$, we know that the solution is given in the form of Eq.~(\ref{explicit}). Moreover, the off-diagonal block matrix $\bf{c}$ turns out to be anti-diagonal with entries equal in modulus but opposite in sign. Therefore, in order to determine if the system has accidentally thermalised, we need only to determine if, at some value of $t$, these entries are identically zero. After some manipulation, this condition reduces to the transcendental equation
\begin{equation}
\label{transcendental}
e^{\gamma_1 t}=\cos\left( 2{\cal J}t \right)-\frac{2{\cal J}}{\gamma_1}\sin\left( 2{\cal J}t \right).
\end{equation}
Interestingly, this `accidental' thermalisation is independent of the temperature of either well and only concerned with the tunnelling strength and the damping rate. For the same parameters taken to obtain the red curve in Fig.~\ref{fidelityopen} {\bf (b)}, we find Eq.~\eqref{transcendental} has two solutions: $t\sim1.03438\omega_1^{-1}$ and $t\sim1.33749\omega_1^{-1}$, clearly corresponding to the two instances of local thermalisation in Fig.~\ref{fidelityopen} {\bf (b)}. Furthermore, we find the thermal occupation numbers of the wells at the first instance of thermalisation are $\nbar_1=1.597$ and $\nbar_2=1.403$, and at the second are $\nbar_1=1.422$ and $\nbar_2=1.578$, thus suggesting that the two instants of accidental thermalisation correspond to an almost swap of the two local thermal states. Increasing $J$ leads to more instances of accidental thermalisation occurring before the system equilibrates to its steady state.\\

\section{Assessment of the non-classical nature of the state of the system.}
\label{quantumness}
Values of fidelity so close to unity should not lead to misinterpretation of the actual nature of the state of the two-well system. In fact, any assessment of fidelity should be accompanied by the study of problem-specific figures of merit able to provide a more fine-grained characterisation of the state at hand. For the sake of a study on thermalisation, a significant class of such quantifiers is embodied by measures of quantum correlations. 

\begin{figure}[b]
{\bf (a)}\\
\includegraphics[width=0.7\columnwidth]{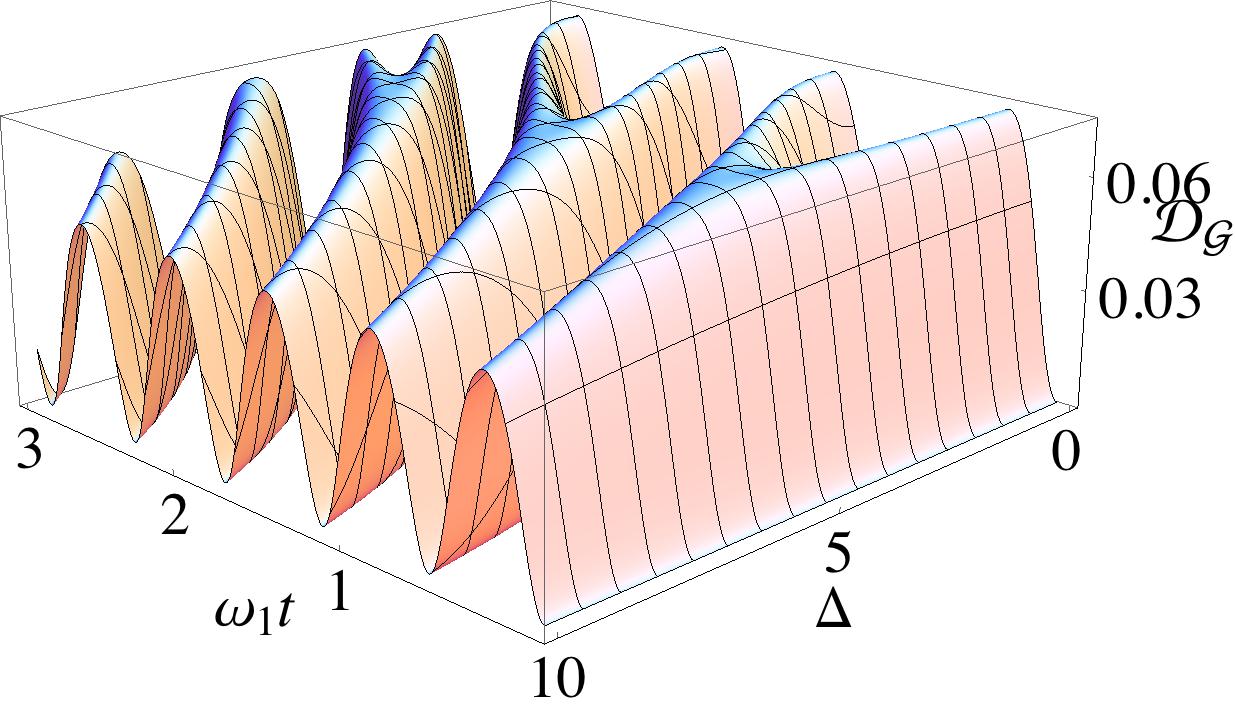}\\
{\bf (b)} \\
\includegraphics[width=0.7\columnwidth]{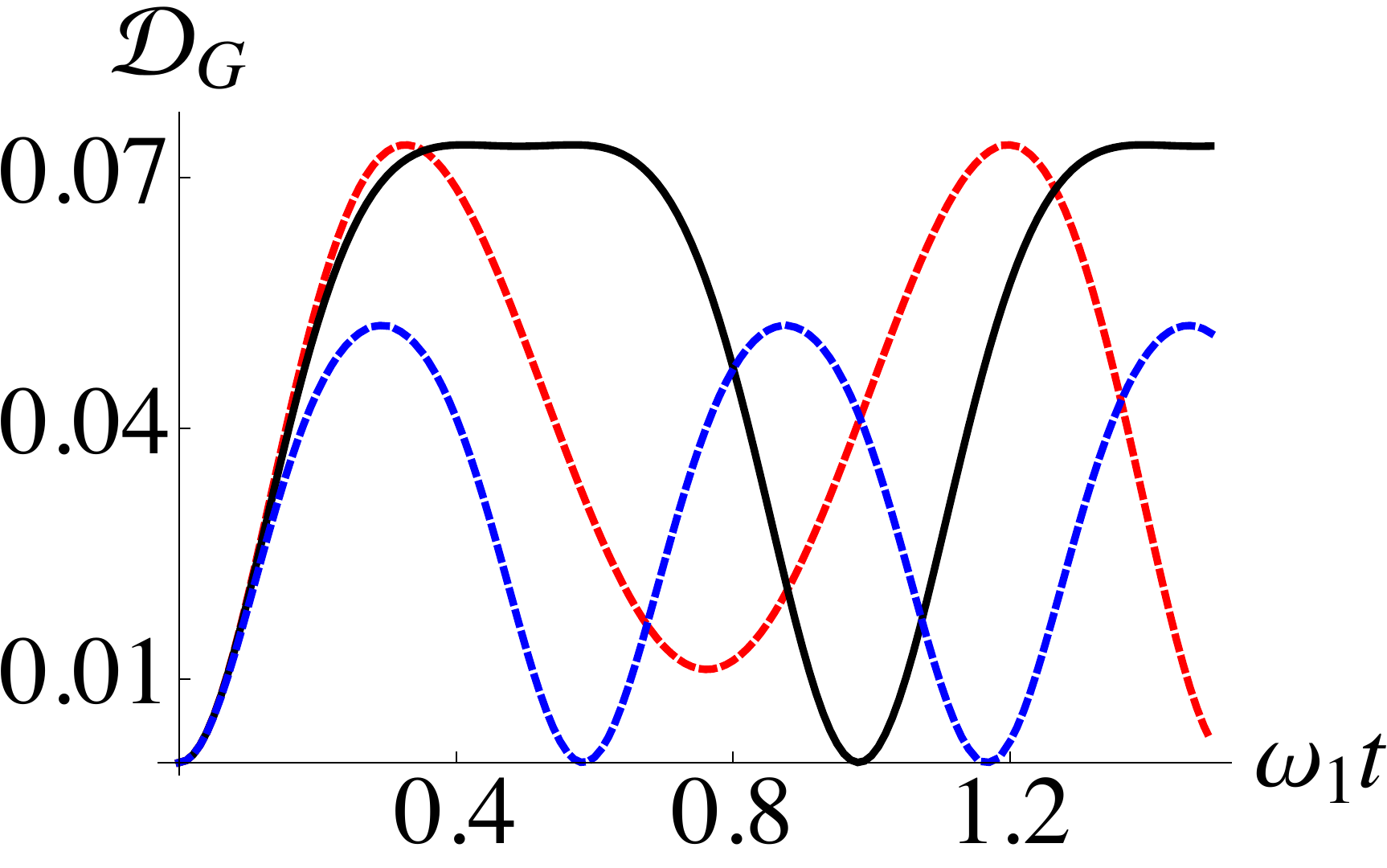}
\caption{{\bf (a)} Discord versus bias and evolution time for the case of closed-system dynamics. We have taken $\nbar_1=1$, $\nbar_2=5$, $J=2$. {\bf (b)} Behaviour of quantum discord in the open-system scenario for $\nbar_1=1$, $\nbar_2=5$, $J=2$ and the bias choices $\Delta=1$ (red curve), 5 (black curve), and 10 (blue curve).}
\label{fig5}
\end{figure}

In this respect, it is important here to assess the role, if any,  various forms of quantum correlations play in the dynamics highlighted above. It is quickly confirmed that, as anticipated before, the system never becomes entangled. While this is expected in light of the nature of the interaction and initial state being considered, nothing prevents the settlement of weaker forms of quantum correlations, such as quantum discord (QD)~\cite{discordreview}. QD is the difference between two classically equivalent definitions of mutual information when applied to a quantum system~\cite{zurek,vedral}. A non-zero degree of QD implies that, in a bipartite system composed of parties A and B, information can be gathered on system A by interrogating party B. For Gaussian states, QD is captured by the Gaussian quantum discord~\cite{adesso,olivares,paris}, which entails that the interrogation of B only involves Gaussian measurements. For a generic covariance matrix $S=\left( \begin{array}{cc}
A & C \\
C^\top & B
\end{array}
\right)$, QD is then defined following Ref.~\cite{adesso} (to be consistent with the definition of the vacuum state used throughout)
\begin{equation}
\mathcal{D}_G = h\left(\sqrt{I_1}\right) - h\left(d_-\right) - h\left(d_+\right) + h\left(\sqrt{E^{\text{min}}}\right),
\label{gaussdisc}
\end{equation}
with 
\begin{widetext}
\begin{equation*}
E^{\text{min}}=\begin{cases}
\dfrac{1}{(I_1-1)^2}\left[{2I_3^2+(I_1-1)(I_4-I_2)+2\vert I_3\vert \sqrt{I_3^2+(I_1-1)(I_4-I_2)}}\right]\quad\text{for}~(I_4-I_1I_2)^2\leq I_3^2(I_2+I_4)(I_1+1), \\~~\\
\dfrac{1}{2I_1}\left[{I_1I_2-I_3^2+I_4-\sqrt{I_3^4+(I_4-I_1I_2)^2-2I_3^2(I_1I_2+I_4)}}\right]\quad\text{otherwise},
\end{cases}
\end{equation*}
\end{widetext}
where
\begin{equation}
\begin{aligned}
&h(x)=\left(\frac{x+1}{2}\right)\log\left(\frac{x+1}{2}\right)-\left(\frac{x-1}{2}\right)\log\left(\frac{x-1}{2}\right),\\ 
&d_\pm^2=\frac{1}{2}\left(\Lambda \pm \sqrt{\Lambda^2-4 I_4}\right),\\
&\Lambda=I_1+I_2+2I_3,
\end{aligned}
\end{equation}
and $I_1=\det A$, $I_2=\det B$, $I_3=\det C$, and $I_4=\det S$. In Fig.~\ref{fig5} {\bf (a)} we study QD against the energy bias for the case of the unitary solution Eq.~\eqref{unitarysol}. Intuitively we would expect that for $\Delta=0$, owing to the full symmetry enforced in the system, QD will be maximised. This is indeed the case, as it can be seen in Fig.~\ref{fig5} {\bf (a)}. However, an interesting feature appears as we increase the bias. At $\Delta/J\sim2.5$, $\nbar_1=1$, and $\nbar_2=5$, QD exhibits a plateau, which implies the existence of an `optimal' value of the bias, dependent on the temperature difference, that helps amplify the non-classicality of the system. Further increase of $\Delta$ pushes the systems too far off resonance, and the coherence decays. In Fig.~\ref{fig5} {\bf (b)} we examine this phenomenon closer, for a fixed temperature difference and small biasing, $\Delta/J=1,$ (red), we see the oscillatory behaviour changes and the first zero-point is lifted. At the optimal value of $\Delta$ (solid black) the plateau is clearly evident. When we increase the bias further, we see the decay in the non-classicality, as well as a change in the periodicity of the system.

\begin{figure}[t!]
{\bf (a)}\\
\includegraphics[width=0.6\columnwidth]{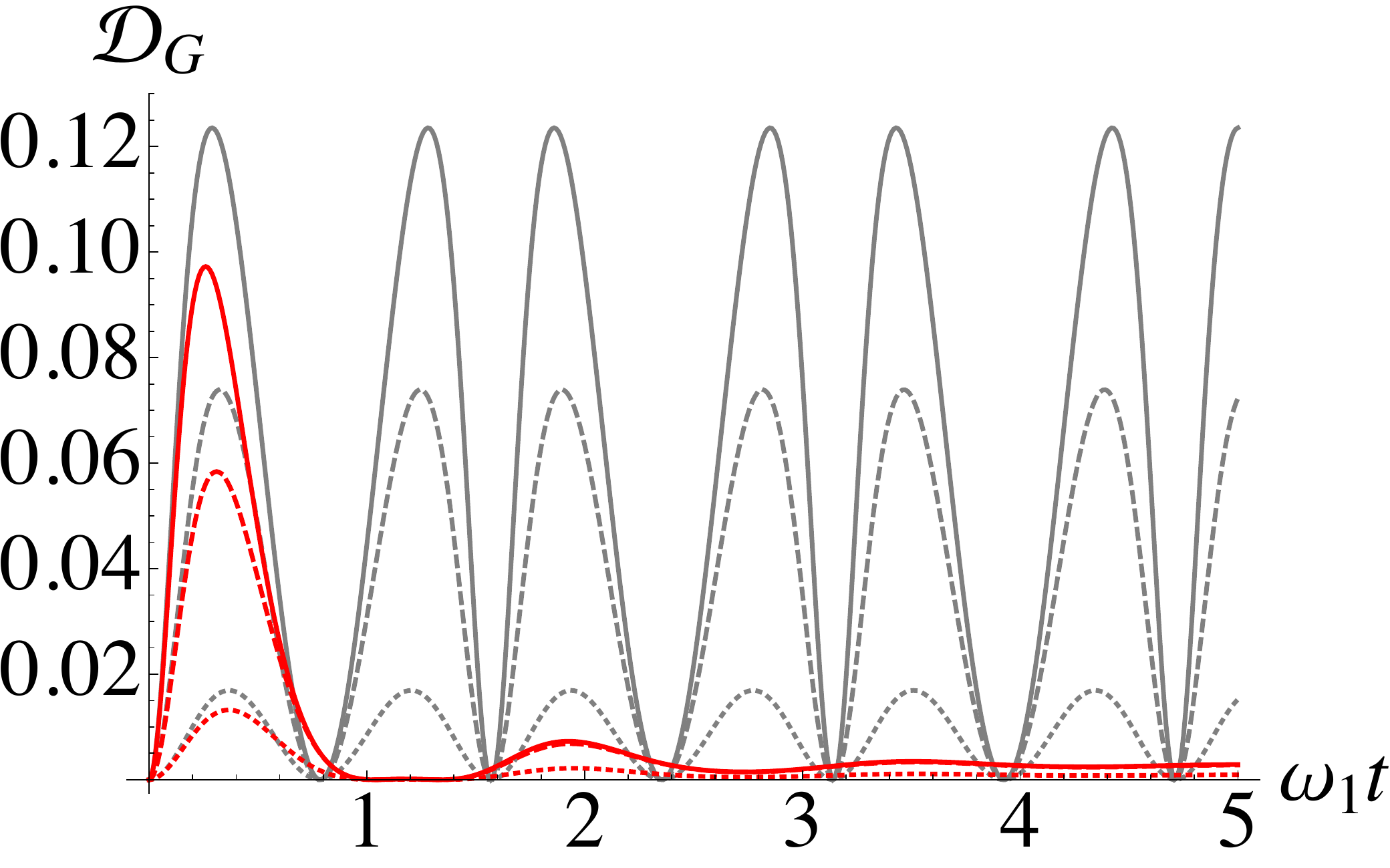}\\
{\bf (b)}\\
\includegraphics[width=0.6\columnwidth]{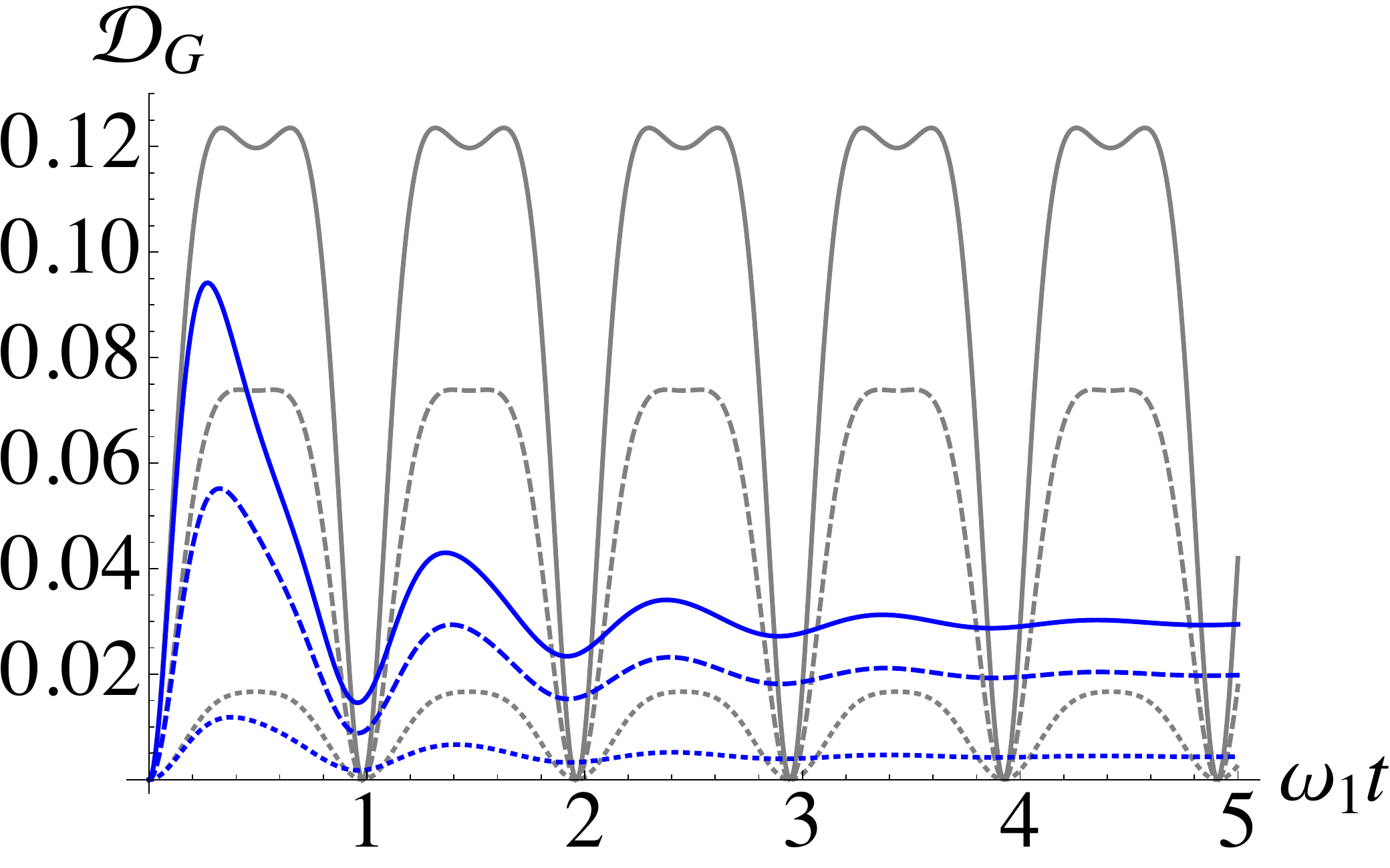}\\
{\bf (c)}\\
\includegraphics[width=0.6\columnwidth]{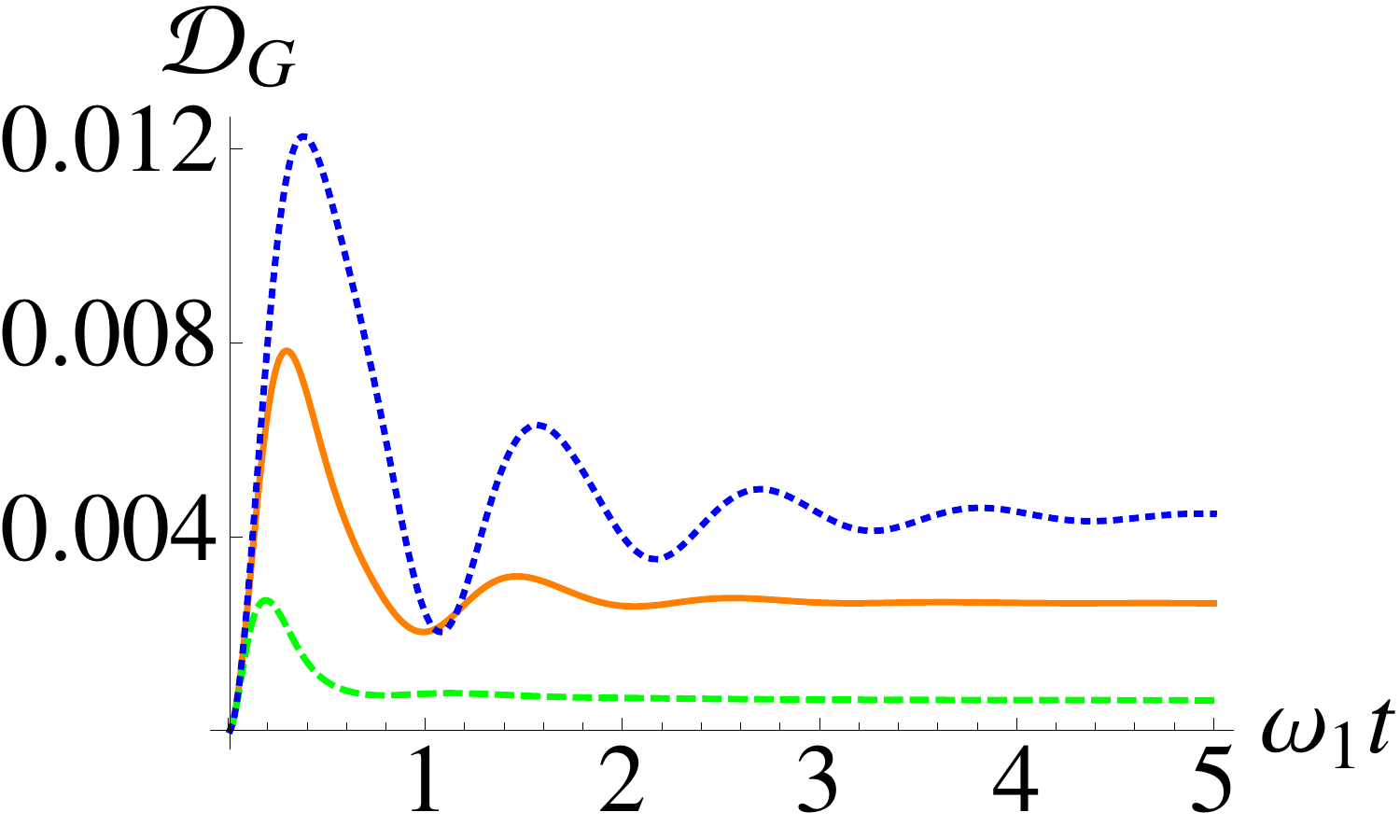}
\caption{{\bf (a)} Dynamical discord for the unitary (gray) and dissipative (red) cases. For both $J=2$, $\nbar_1=1$, and $\nbar_2=$ 2 (dotted), 5 (dashed), and 10 (solid). For all the dissipative cases $\gamma_1/\omega_1=\gamma_2/\omega_1=1$. {\bf (b)} As for panel {\bf (a)} but with the optimal bias (for $\nbar_1=1$, $\nbar_2=5$) between the wells, $\Delta=5$.  {\bf (c)} Dissipative dynamical discord $J=2$, $\Delta=4$, $\gamma_1/\omega_1=\gamma_2/\omega_1=1$, $\nbar_1=1$, and $\nbar_2=2$, with self interaction term $U=0$, $U=1$ and $U=3$ going from top to bottom.}
\label{fig6}
\end{figure}

Turning our attention to the dissipative case, in Fig.~\ref{fig6} we compare the unitary dynamics with the dissipative case for unbiased wells [panel {\bf (a)}], and biased ones [panel {\bf (b)}] at various differences in temperature. For unbiased wells, we see dissipation quickly suppresses the the oscillations and we reach a steady state with non-vanishing QD. As we increase the temperature between the wells we see an increase in the QD for the unitary case and the steady state QD is larger for increasing temperature difference. When we bias the wells, taking $\Delta/J=2.5$ for all temperature differences, we see the dissipative dynamics clearly show the enhanced non-classicality. While the time to reach equilibrium appears unaffected, the steady state is significantly more non-classical than in the unbiased situation. This may imply that in this situation the non-classicality plays no role in reaching equilibrium. In Fig.~\ref{fig6} {\bf (c)} we examine the effect that self-interaction has on the dynamics of nonclassicality. In order to do so, we compute the Gaussian discord of the hypothetical Gaussian state having, as covariance matrix, the one achieved by calculating the entries $\sigma_{ij}$ over the non-Gaussian state resulting from a chosen non-zero values of $U$. Evidently the larger the self interaction, the more self ordered each well becomes, diminishing the effect of the tunnelling and reducing the amount of nonclassicality present. Also we see the system tends to equilibrate faster.

\begin{figure}[t]
{\bf (a)} \hskip0.45\columnwidth {\bf (b)}\\
\includegraphics[width=0.45\columnwidth]{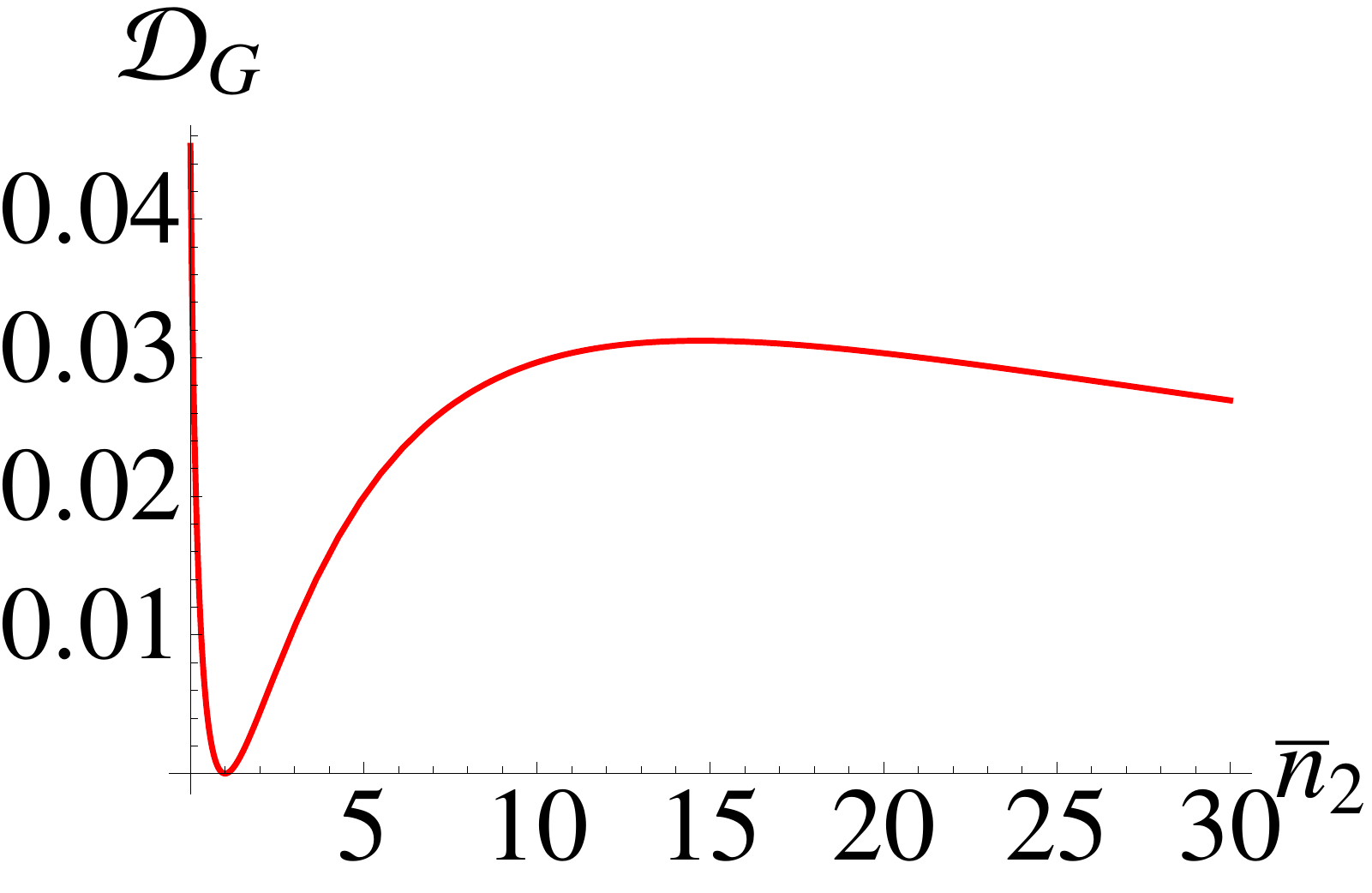}~\includegraphics[width=0.45\columnwidth]{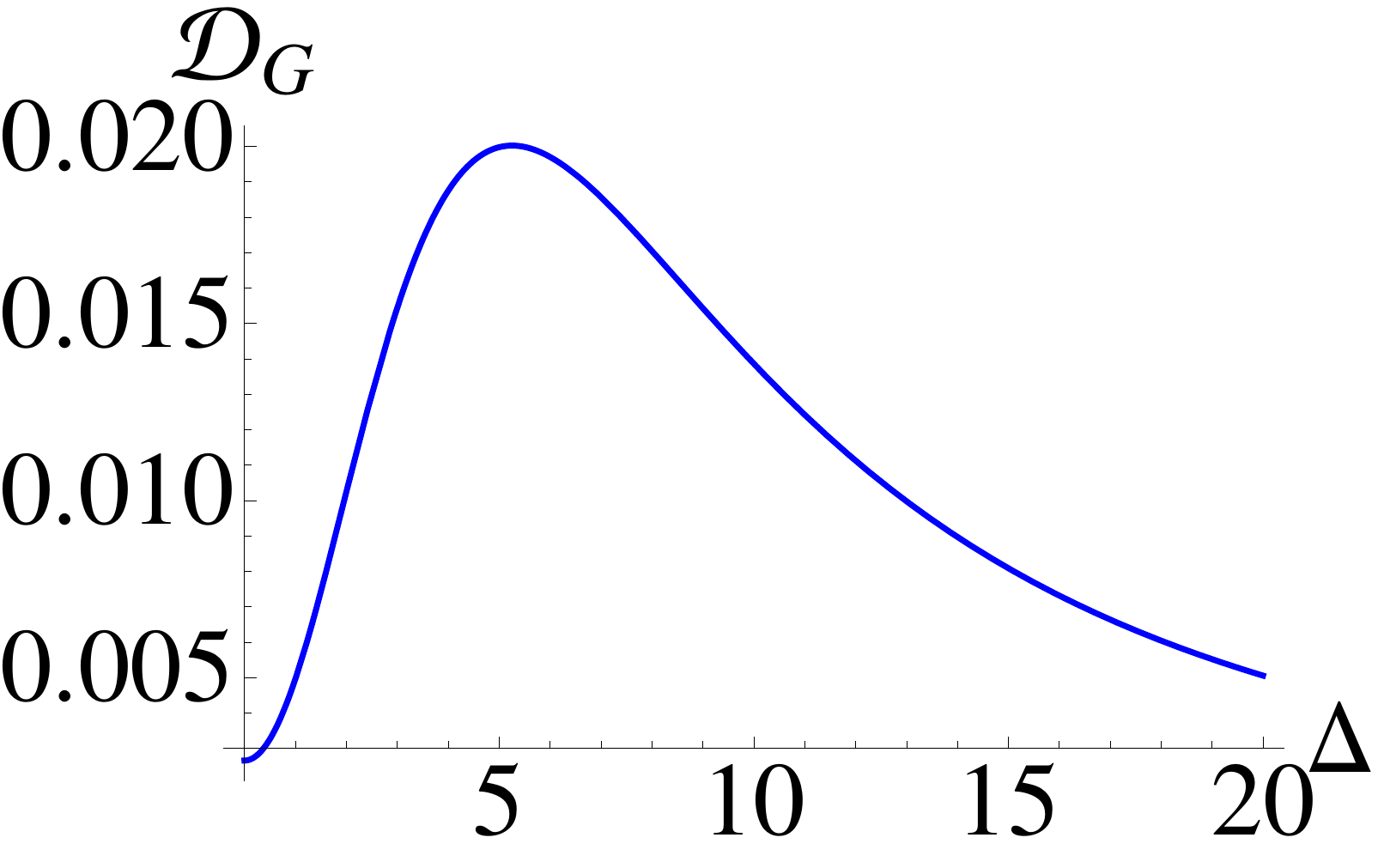}\\
{\bf (c)} \hskip0.45\columnwidth {\bf (d)}\\
\includegraphics[width=0.45\columnwidth]{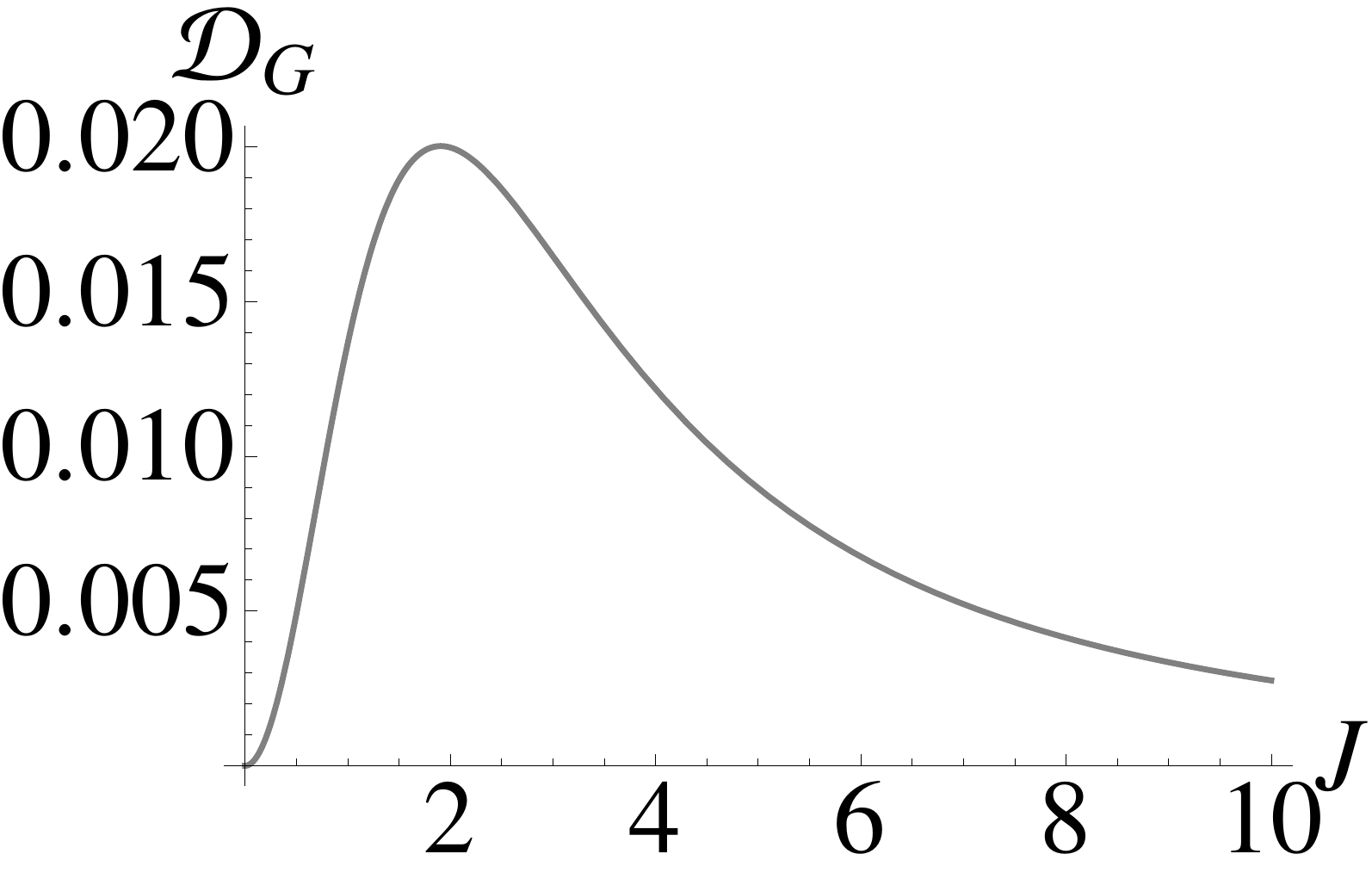}~\includegraphics[width=0.45\columnwidth]{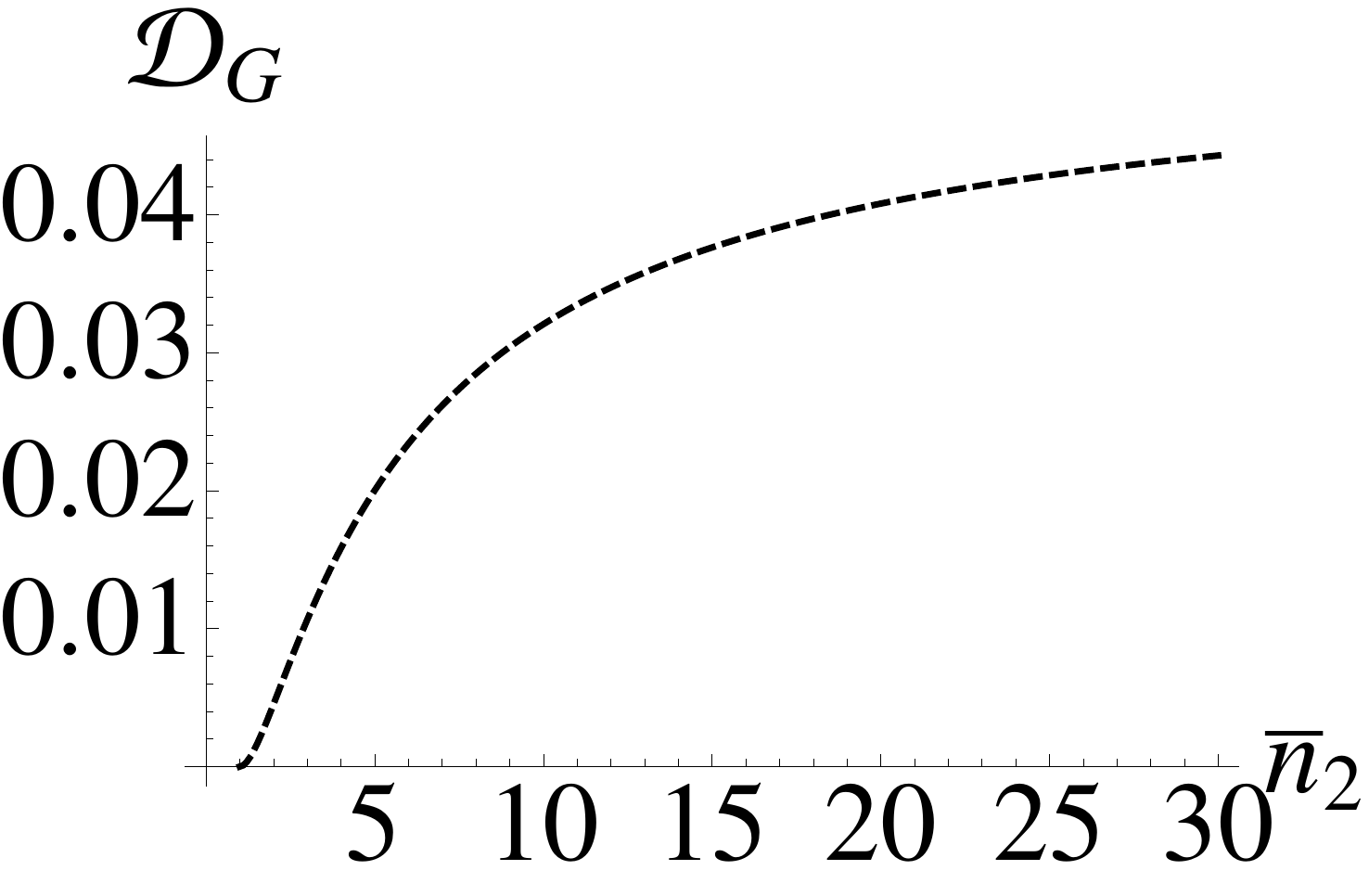}\\
{\bf (e)} \hskip0.45\columnwidth {\bf (f)}\\
\includegraphics[width=0.45\columnwidth]{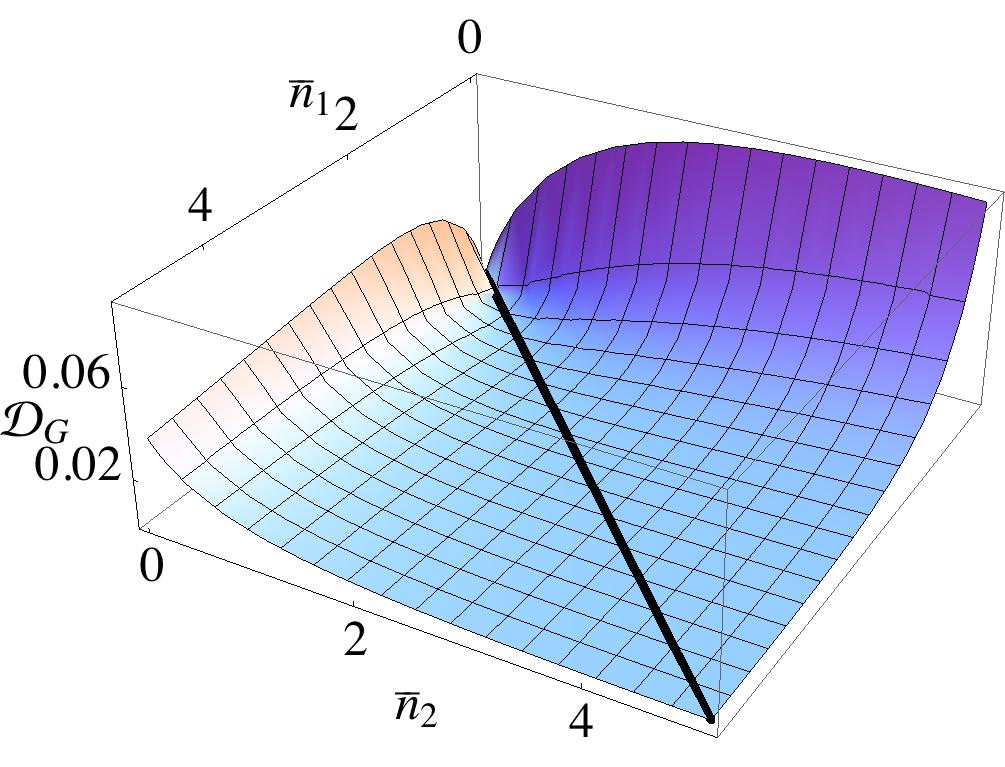}~\includegraphics[width=0.45\columnwidth]{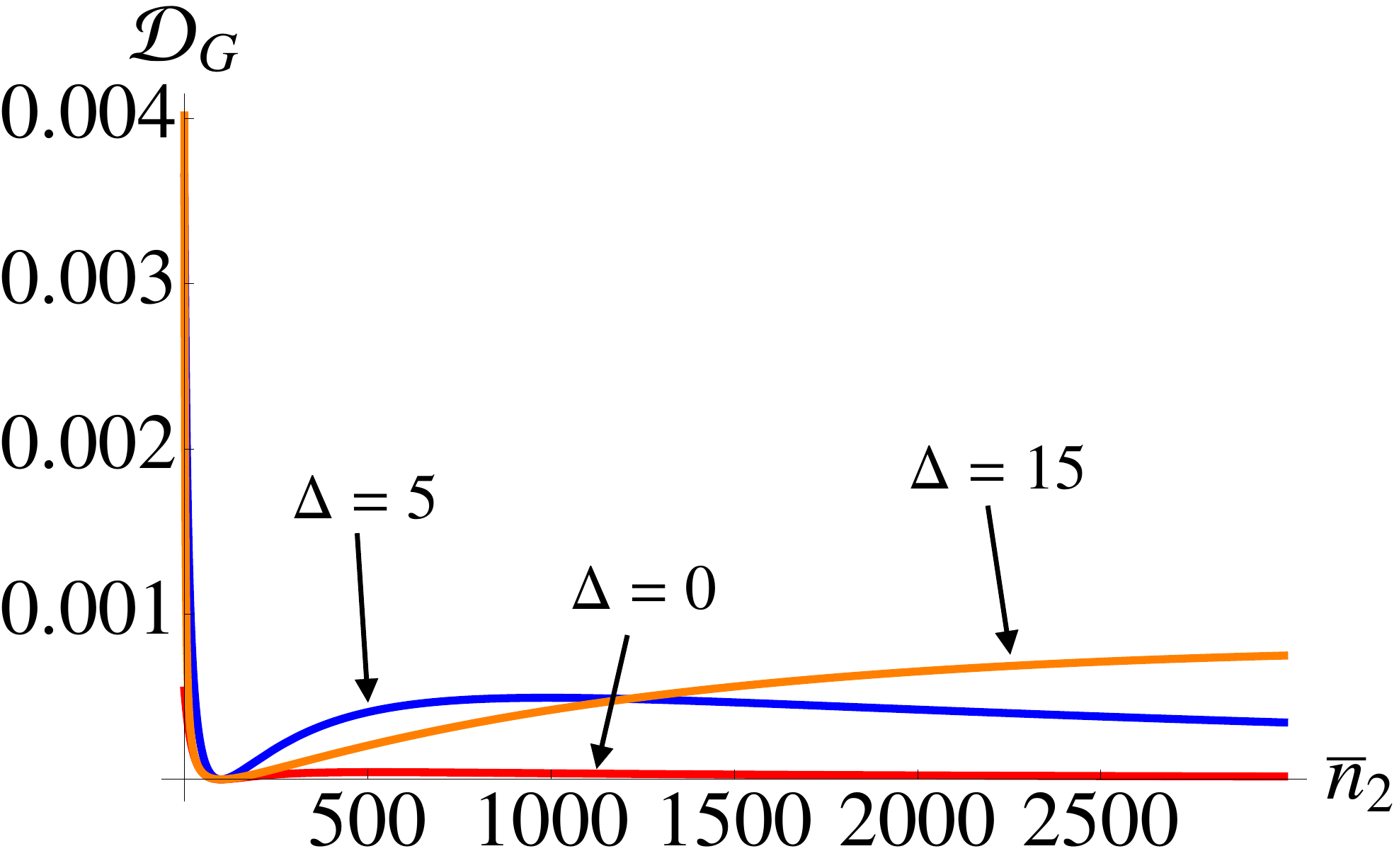}
\caption{Steady-state discord between the two wells for $\gamma_1/\omega_1=\gamma_2/\omega_1=1$, and $\nbar_1=1$. {\bf (a)} Plotted against $\nbar_2$ for $J=2$, $\Delta=5$. {\bf (b)} Against $\Delta$ for $J=2$ and $\nbar_2=5$. {\bf (c)} Against $J$ for $\nbar_2=5$ and $\Delta=5$. {\bf (d)} Maximum discord attainable for a given value of $\nbar_2$ found by optimising with respect to both $\Delta$ and $J$ when $\nbar_1=1$ and $\gamma_1/\omega_1=\gamma_2/\omega_1=1$. {\bf (e)} Steady-state discord studied against both $\nbar_1$ and $\nbar_2$ when $J=2$, $\Delta=5$, and $\gamma_1/\omega_1=\gamma_2/\omega_1=1$. The black line shows that $\mathcal{D}_G$ is identically null only when $\nbar_1=\nbar_2$. {\bf (f)} Steady-state discord against $\nbar_2$ for $\nbar_1=100$ with $J=2$, $\gamma_1/\omega_1=\gamma_2/\omega_1=1$.}
\label{fig7}
\end{figure}

The nonclassicality of the steady state is delicately dependent on the temperature difference, as well as the tunneling rate and the bias. In Fig.~\ref{fig7} we examine this behaviour closer, fixing $\nbar_1$ and $\gamma_1/\omega_1=\gamma_2/\omega_1=1$ with $J=2$ and $\Delta=5$. The only conditions for which the system does not exhibit nonclassical correlations is the trivial one of $\nbar_2=\nbar_1$. As we increase the temperature imbalance we see that QD increases to a maximum value before slowly decaying [cf. Figs.~\ref{fig7} {\bf (a)} and {\bf (e)}]. If we fix the temperature difference such that $\nbar_1=1$ and $\nbar_2=5$, we see in panels {\bf (b)} and {\bf (c)} that there are optimal values of the remaining parameters that give the largest value of discord. While the reservoirs have been kept at moderately low energies, in panel {\bf (f)} we significantly increase both $\nbar_1$ and $\nbar_2$ and see that large values of QD can still be achieved. An unbiased configuration leads to values of discord of the order of $10^{-5}$. Increasing the bias, such values are raised by  up to one order of magnitude.\\

\section{Dynamics of the energy flux between the wells} 
It is important to gather insight into the details of the exchange of energy between the wells of the system, which is at the basis of the process of quasi-thermalisation highlighted so far and takes place in two forms: an exchange of particles between the wells and a similar process occurring at the interface between the double-well system and the reservoirs. The aim of this section is to identify the contribution coming from both such fluxes. We are thus interested in quantifying the flux into/from well $j=1,2$, which we label as $\dot{\mathcal{Q}}_j$
, and the total flux $\dot{\mathcal{Q}}_{tot}$. These are given by the quantities
\begin{equation}
\label{fluxes}
\dot{\mathcal{Q}}_{tot} = \text{Tr}[\hat{\cal H} \partial_t \varrho],\qquad\dot{\mathcal{Q}}_j = \text{Tr}[\hat{\cal H}_j \partial_t \varrho_j], ~~(j=1,2),\\
\end{equation}
where $\hat{\cal H}_j=\omega_j(\hat{a}^\dag_j\hat a_j+1/2)$ is the free evolution of a single well and $\varrho_j$ is the density matrix of well $j$. Conveniently, these quantities can be directly evaluated from the covariance matrix (and we will assume both wells to have the same damping rate, i.e. $\gamma_1/\omega_1=\gamma_2/\omega_1=\gamma$). We find 
\begin{equation}
\begin{aligned}
&\dot{\mathcal{Q}}_1 =  e^{-\gamma t} \left[ \frac{2J^2(1+\Delta)(\nbar_2-\nbar_1)}{\sqrt{4 J^2+\Delta^2}} \sin\left(\sqrt{4J^2+\Delta^2} t\right)  \right],\\
&\dot{\mathcal{Q}}_2 =-(1+\Delta)\dot{\mathcal{Q}}_1,\qquad\dot{\mathcal{Q}}_{tot} = 0.
\end{aligned}
\end{equation}
From here it is easy to confirm that ${\dot{\mathcal{Q}}_1}/{\omega_1} = - {\dot{\mathcal{Q}}_2}/{\omega_2}$ and clearly taking $\gamma=0$ or $\Delta=0$ recovers the unitary and unbiased limits respectively. However, this behaviour is only for the special case of the wells initially being thermalised with their baths, while taking a different initial state this behaviour no longer holds. Indeed, what is special about our initial state is that it conserves the total energy of the system. This is readily seen given that $\dot{\mathcal{Q}}_{tot}=0$ and it is easy to confirm that
\begin{equation}
\mathcal{Q}_{tot}=1+\frac{\Delta}{2}+\nbar_1+(1+\Delta)\nbar_2,
\end{equation}
for all $t$ and $J$. Of course, the energy of the individual wells changes dynamically (until settling into the same steady state). 

We can gain further insight into the reason for this by examining closer the quantity we are calculating, i.e.
\begin{equation*}
\begin{aligned}
\dot{\mathcal{Q}}_{tot}&=\text{Tr}[\hat{\mathcal{H}} \partial_t\varrho]\\
&=\text{Tr}\left[\hat{\mathcal{H}} \left(-i[\hat{\mathcal{H}},\varrho] + \sum_{i=1}^2 \mathcal{L}_i(\varrho)   \right) \right]\\
&=-i\text{Tr}\left[ \hat{\mathcal{H}}\left(\hat{\mathcal{H}}\varrho-\varrho\hat{\mathcal{H}}\right) \right] + \text{Tr}\left[  \hat{\mathcal{H}} \sum_{i=1}^2 \mathcal{L}_i(\varrho)  \right]\\
&=\text{Tr}\left[  \hat{\mathcal{H}} \sum_{i=1}^2 \mathcal{L}_i(\varrho)  \right].
\end{aligned}
\end{equation*}
The tunnelling term in Eq.~\eqref{hamiltonian} commutes with $\mathcal{L}_i$, and when $U=0$ the only contribution to the total flux is from the free evolution of each well. Therefore we are interested in calculating
\begin{equation}
\label{expression}
\dot{\mathcal{Q}}_{tot}=\text{Tr}\left[ \left( \nbar_1+\nbar_2  \right) \mathcal{L}_1(\varrho) \right]+\text{Tr}\left[ \left( \nbar_1+\nbar_2  \right) \mathcal{L}_2(\varrho) \right].
\end{equation}
In a tedious but otherwise straightforward calculation, we can explicitly evaluate this expression when assuming the special initial condition $\varrho=\frac{e^{-\beta_1 \hat{\cal H}_1}}{\mathcal{Z}_1}\otimes\frac{e^{-\beta_2 \hat{\cal H}_2}}{\mathcal{Z}_2}$, with ${\cal Z}_j=\text{Tr}[e^{-\beta_j \hat{\cal H}_j}]$. We find that both the terms entering Eq.~(\ref{expression}) are identically zero, thus showing that, in the $U=0$ case, the net heat flux is null due to two special circumstances: on one hand our chosen initial state, on the other hand the tunnelling term commutes with the super operators. 

\begin{figure}[t]
{\bf (a)}\\
\includegraphics[width=0.6\columnwidth]{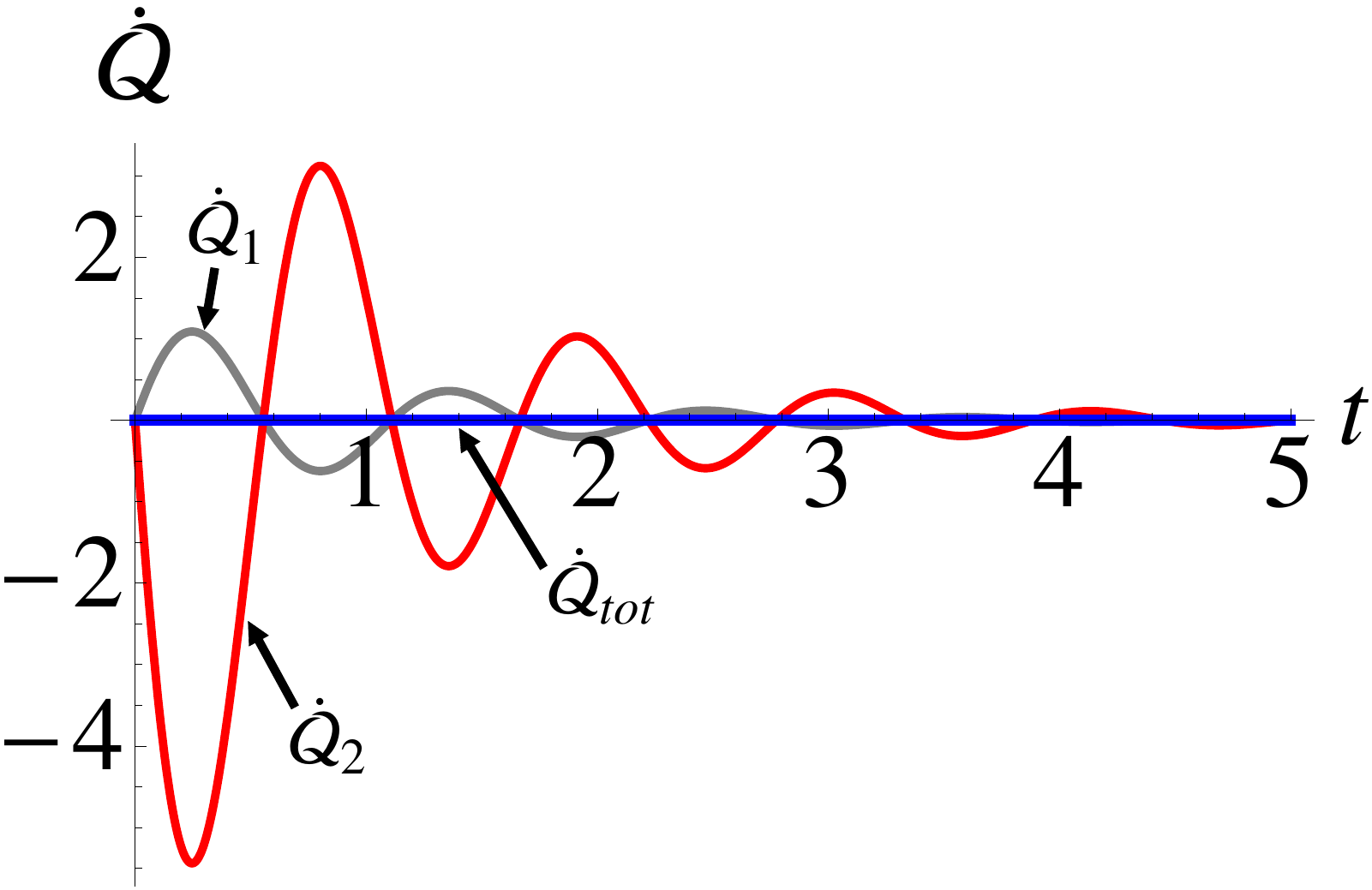}\\
{\bf (b)}\\
\includegraphics[width=0.6\columnwidth]{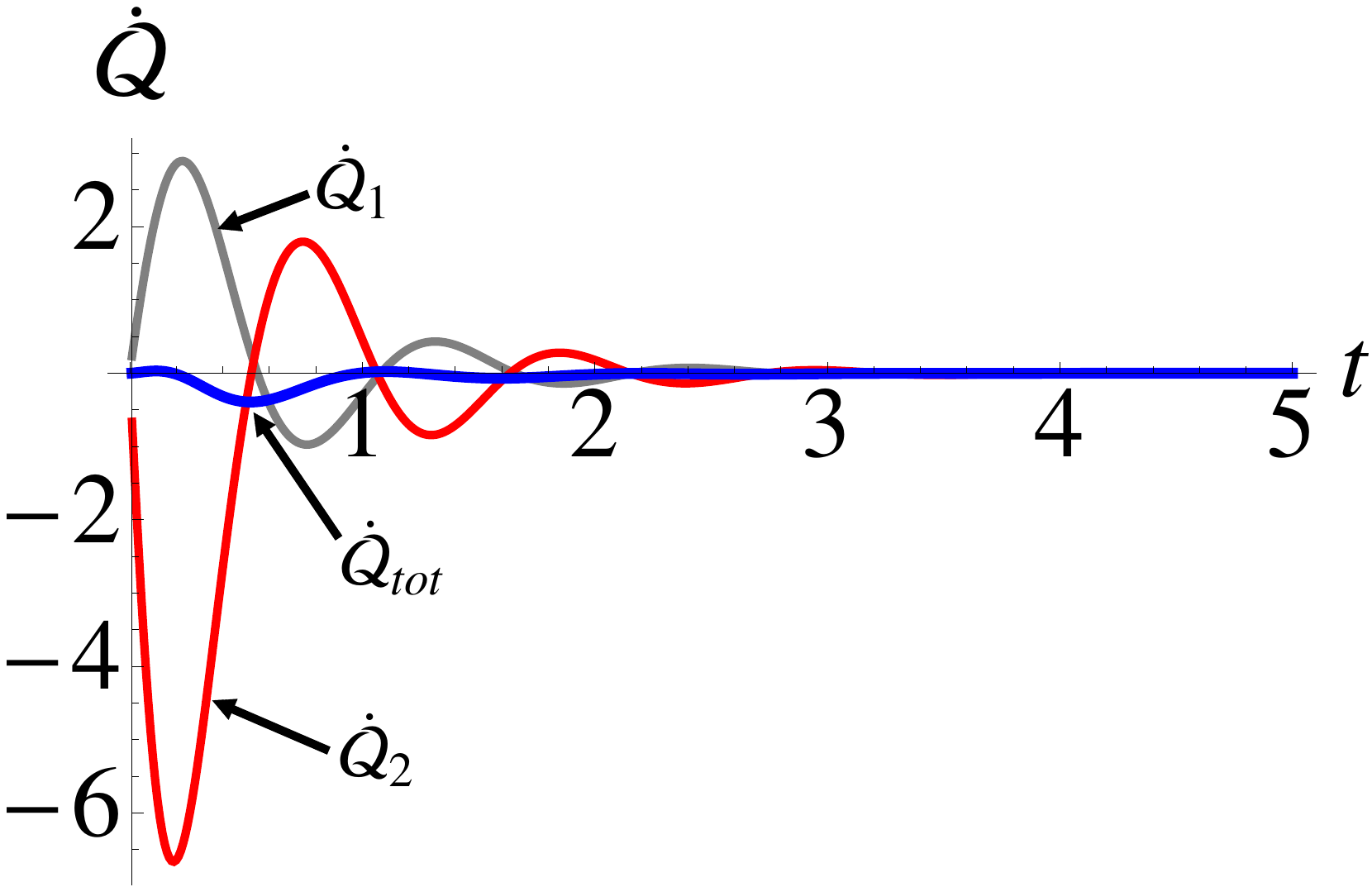}\\
{\bf (c)}\\
\includegraphics[width=0.6\columnwidth]{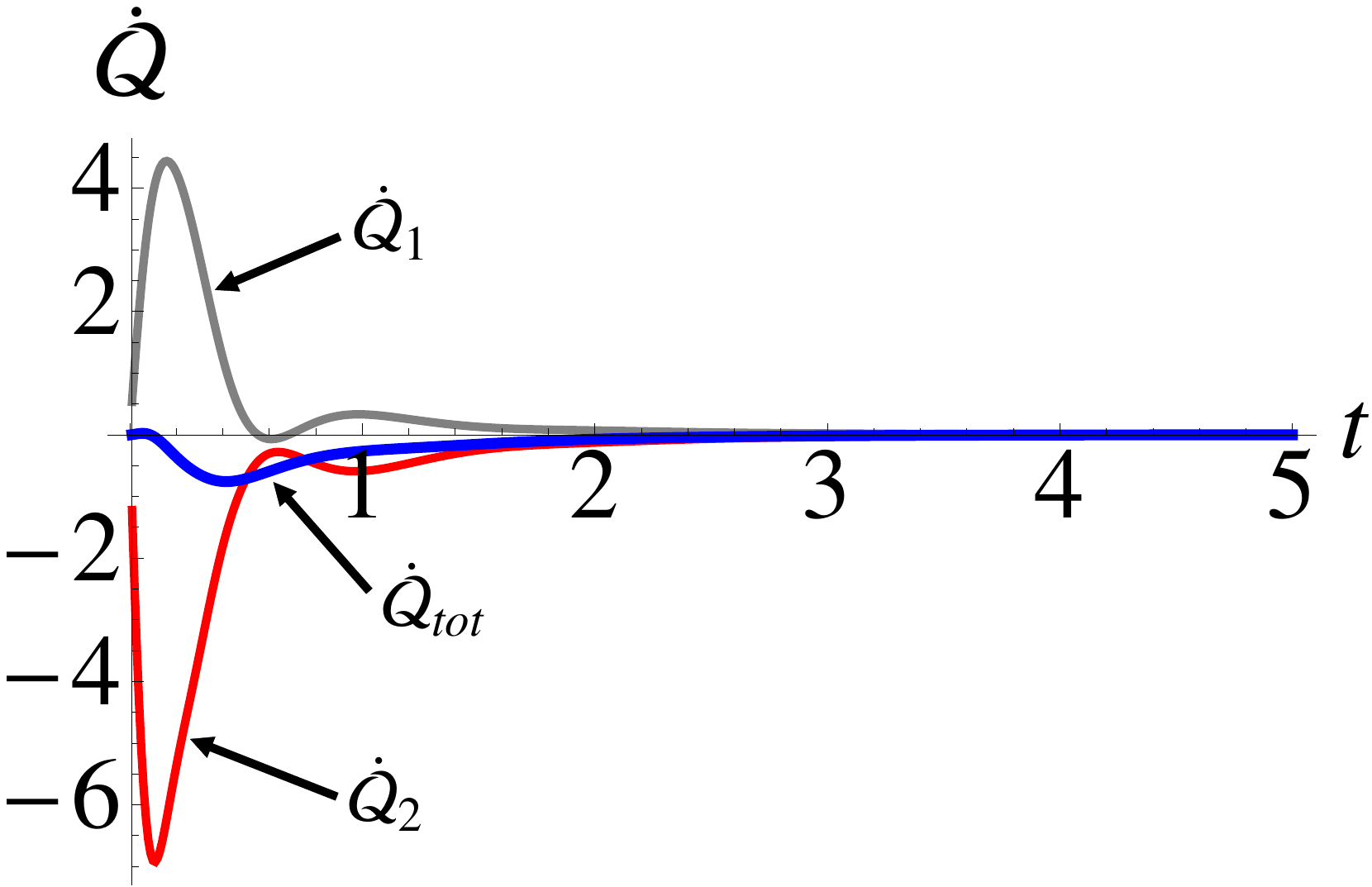}
\caption{Steady-state discord between the two wells. In all panels $J=2,~\gamma=1,~\nbar_1=1,~\nbar_2=2,~\Delta=4.$ Red Line: hotter well (well 2), gray line cooler well (well 1), and blue the total flux. With {\bf (a)} $U=0.$ {\bf (b)} $U=1.$ {\bf (c)} $U=3.$}
\label{fig8}
\end{figure}

In Fig.~\ref{fig8} {\bf (a)} we show the dynamics of the various energy fluxes given in Eq.~\eqref{fluxes}. We notice how the flux into the cooler well is proportional to the flux out of the hotter well, which results in a null net flux. Needless to say, the single-well fluxes only account for the net intake/outtake of particles for one of the wells and do not provide information on the actual balance between the contribution due to the coupling to the reservoir and that due to the coherent inter-well interaction.

We can study the intermediate dynamical regime where the self-interaction is non-zero and comparable with the tunnelling by numerically solving Eq.~\eqref{master} and examining the behaviour of the heat fluxes, of which we illustrate some examples in Fig.~\ref{fig8} {\bf (b)} and {\bf (c)} [we refer to the caption for an account of the parameters used in the simulations]. The total flux is now non-zero, and the energy is not conserved. However, the average occupation number is conserved, i.e. ${\langle \hat{a}_1^{\dagger}\hat{a}_1 \rangle}/{\omega_1} = -{\langle \hat{a}_2^{\dagger}\hat{a}_2 \rangle}/{\omega_2}$, which follows directly from the previous arguments. \\

\section{Discussion}
The analysis above shows that neither global nor local thermalisation with the reservoirs is achieved. The fidelity between the density matrices of the time-evolved state and the target thermal one (whether globally or locally) connects the closeness of the populations of the energy levels of the former to the statistics of the latter. However, the interaction between the wells establishes strong quantum coherence between the particles of the systems, which in turn results in the generation of a substantive degree of quantum correlations, albeit of a nature weaker than entanglement, which prevents the thermal character of the resulting state. 

The analysis reported here also has the merit of providing rather deep insight into the phenomenology of quantum correlations between the wells. We have qualitatively and quantitatively examined the dynamics and steady state of a BEC loaded into a double well potential. While the wells remain separable at all times, thus sharing no entanglement, by exploring the behaviour of the quantum discord we find the system to be always non-classical, except under trivial, uninteresting conditions. Furthermore, the degree of nonclassicality of the system is dependent on the energy bias between the two wells. For identical wells, a significant amount of QD is possible, provided that a large temperature imbalance is established. Such nonclassicality can be greatly enhanced by taking a suitable value of tunnelling, which must be a function of the given bias. The transfer of heat in the system is equally complex. \\

\begin{widetext}
\section{Methods}
{\bf Differential Equations.} 
Here we provide the complete set of differential equations that describe the dissipative dynamics considered throughout.
\begin{equation}
\label{diffs}
\begin{matrix}
\partial_t y_1 = -J z_2 -\frac{\gamma_1}{2} y_1 + \omega_1 z_1,\quad\partial_t y_2 = -J z_1 -\frac{\gamma_2}{2} y_2 + \omega_2 z_2,&\partial_t z_1 = J y_2 - \frac{\gamma_1}{2} z_1 - \omega_1 y_1,\quad\partial_t z_2 = J y_1 - \frac{\gamma_2}{2} z_2 - \omega_2 y_2,\\
\partial_t \sigma_1^x = \gamma_1 +2\nbar_1 \gamma_1 - 2 J \sigma_{12}^{xp} -\gamma_1 \sigma_1^x + 2\omega_1 \sigma_1^{xp},
&\partial_t \sigma_2^x = \gamma_2 +2\nbar_2 \gamma_2 - 2 J \sigma_{12}^{px} -\gamma_2 \sigma_2^x + 2\omega_2 \sigma_2^{xp}, \\
\partial_t \sigma_1^p=  \gamma_1 +2\nbar_1 \gamma_1 + 2 J \sigma_{12}^{px} -\gamma_1 \sigma_1^p - 2\omega_1 \sigma_1^{xp},
&\partial_t \sigma_2^p=  \gamma_2 +2\nbar_2 \gamma_2 + 2 J \sigma_{12}^{xp} -\gamma_2 \sigma_2^p - 2\omega_2 \sigma_2^{xp},  \\
\partial_t\sigma_1^{xp}=J\left( \sigma_{12}^x-\sigma_{12}^p \right) - \gamma_1 \sigma_1^{xp} - \omega_1 \left( \sigma_1^x - \sigma_1^p \right),
&\partial_t\sigma_2^{xp}=J\left( \sigma_{12}^x-\sigma_{12}^p \right) - \gamma_2 \sigma_2^{xp} - \omega_1 \left( \sigma_2^x - \sigma_2^p \right),   \\
\partial_t\sigma_{12}^x=-J\left( \sigma_1^{xp} + \sigma_2^{xp} \right) - \frac{\gamma_1+\gamma_2}{2} \sigma_{12}^x + \omega_1 \sigma_{12}^{px} + \omega_2\sigma_{12}^{xp},
&\partial_t\sigma_{12}^p=J\left( \sigma_1^{xp} + \sigma_2^{xp} \right) - \frac{\gamma_1+\gamma_2}{2} \sigma_{12}^p - \omega_1 \sigma_{12}^{xp} -  \omega_2\sigma_{12}^{px}, \\
\partial_t\sigma_{12}^{xp}=J\left( \sigma_1^{x} - \sigma_2^{p} \right) - \frac{\gamma_1+\gamma_2}{2} \sigma_{12}^{xp} + \omega_1\sigma_{12}^p -  \omega_2 \sigma_{12}^x,
&\partial_t\sigma_{12}^{px}=J\left( \sigma_2^{x} - \sigma_1^{p} \right) - \frac{\gamma_1+\gamma_2}{2} \sigma_{12}^{px} - \omega_1\sigma_{12}^x +  \omega_2 \sigma_{12}^p.
\end{matrix}
\end{equation}
\end{widetext}

\noindent
{\bf Discussions on self interaction dominated limit.} 
While the main analysis treated the tunnelling dominated regime, the opposite extreme is determined setting $J=0$ and exploring the situation where self-interaction dominates. In this instance, the two wells are completely decoupled from one another. We can directly solve Eq.~\eqref{master} by projecting onto the number states $\ket{n}$. Since these states are eigenstates of the Hamiltonian for $J=0$ the steady-state will be entirely independent of $U$. In fact, regardless of the initial state we find the steady state for each well to be $\rho_j=\frac{1}{\mathcal{Z}_j}e^{-\beta_j \omega_j \hat{n}_j}$ with $e^{-\beta_j \omega_j}=\frac{\nbar_j}{\nbar_j+1}$, which is the Boltzmann distribution for a harmonic oscillator with thermal occupation $\nbar_j$. Clearly then, if our initial states are already thermalised with their local reservoir we see no dynamics. For any other initial state, the two wells thermalise independently to their respective reservoir temperatures.\\

\acknowledgements
SC is grateful to Simon Pigeon, Lorenzo Fusco, and Alessandro Ferraro for helpful and enlightening discussions. The authors acknowledge support from the UK EPSRC (EP/L005026/1, EP/M003019/1), the John Templeton Foundation (grant ID 43467), and the EU Collaborative Project TherMiQ (Grant Agreement 618074).

\end{document}